\documentclass[aps,twocolumn,floatfix,rmp]{revtex4}
\usepackage{amsmath,amssymb,graphicx}
\newcommand{\etal}{{\it et al.}}

\begin{document}

\title{Unconventional Superconductivity} 

\author{M. R. Norman}
\affiliation{Materials Science Division, Argonne National Laboratory, Argonne IL 60439}

\begin{abstract}
A brief review of unconventional superconductivity is given, stretching from the halcyon days of helium-3
to the modern world of Majorana fermions.  Along the way, we will encounter such strange beasts 
as heavy fermion superconductors, cuprates, and their iron-based cousins.  Emphasis will be put 
on the fact that in almost all cases, an accepted microscopic theory has yet to emerge.  This is 
attributed to the difficulty of constructing a theory of superconductivity outside the Migdal-Eliashberg 
framework.
\end{abstract}
\date{\today}

\maketitle

\tableofcontents

\section{Introduction to Unconventional Superconductors}

Superconductivity was discovered in 1911 in an attempt to understand how the resistivity of a metal
behaved at low temperatures \cite{HK}.  The ideas on the table were that the resistance would monotonically
go to zero as absolute zero was approached, it would saturate, or it would diverge.  One can imagine
the surprise of Onnes' group when instead, the resistivity in mercury plummeted to zero at a particular
temperature, T$_c$.  Many famous theorists, including Einstein and Heisenberg, attempted to elucidate
its origin, but it took the development of modern many-body theory in the 1950s before a proper toolkit 
emerged for its solution.  Still, having the tools and coming up with a solution were two quite different
things.  It was the remarkable insight of John Bardeen coupled with the talents of a young postdoc,
Leon Cooper, and an even younger student, Bob Schrieffer, that led to its ultimate solution \cite{bcs}, in the
process beating out such luminaries as Lev Landau and Richard Feynman.

At the heart of the so-called BCS theory is the concept of Cooper pairs \cite{cooper}.  What Cooper found
was that an arbitrarily weak attractive interaction between electrons would lead to a profound rearrangement
of the Fermi surface, leading to the formation of quasi-bound electron pairs.  At a fell swoop, this solved
many of the outstanding issues of superconductivity, particularly the existence of an energy gap.
And, unlike fermions which typically do not condense, pairs of fermions, being statistically equivalent to
bosons, can condense, which in turn can lead to a zero resistance state as well as to the famous Meissner
effect \cite{meissner,london}, where magnetic flux is expelled from a superconductor when going below T$_c$.

But Cooper pairs are very different from the real space pairs that had been suggested by  
Schafroth \cite{blatt}.  Cooper pairs are constructed in momentum space, where one correlates an electron at ${\bf k}$
with its time reversed partner at $-{\bf k}$.  In real space, these correlations extend out to a distance known
as the coherence length which is typically much larger than the inter-particle separation.  In momentum space,
these correlations occur in an energy shell about the Fermi surface, very different from the Bose-Einstein condensation
limit of real space pairs where the chemical potential is well below the bottom of the fermionic band.

\begin{figure}
\includegraphics[width=1.5in]{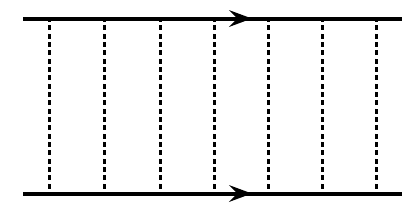}
\caption{Particle-particle ladder sum that gives rise to the Cooper instability.  Solid lines are electrons, dashed lines the
pair interaction.}
\label{ladder}
\end{figure}

Although the BCS theory is one of the most profound many body theories ever discovered in science, it is at heart
a weak coupling mean field theory.  Its great success followed two subsequent developments.  The first was the 
realization by Gor'kov that the theory was equivalent to the more general Ginzburg-Landau theory based on a phenomenological 
order parameter \cite{gorkov}.  This opened up a large vista of applications, since the simplicity of that theory could
be applied to a large variety of problems, including the spatial variation of the order parameter \cite{deGennes}.  The second
was the realization by Migdal \cite{migdal} that the success of BCS theory was based on the fact that a controlled
perturbation expansion existed for the electron-ion interaction that was at the heart of the BCS mechanism.  In
essence, besides the repeated scattering of electrons that leads to the infrared Cooper singularity (the ladder sum shown
in Fig.~\ref{ladder}), all other Feynman diagrams are controlled by an expansion in the small parameter $\hbar \omega_D/E_F$,
where $\omega_D$ is the Debye frequency of the ions, and $E_F$ the Fermi energy of the electrons.  This led to
the generalization of the BCS theory by Nambu \cite{nambu} and Eliashberg \cite{grisha} to take into account the
frequency dependence of the normal and anomalous (pairing) self-energies.

\begin{figure}
\includegraphics[width=1.75in]{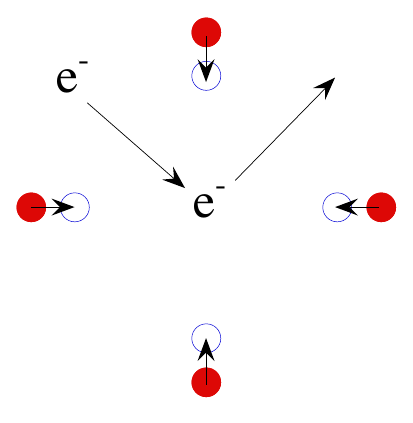}
\caption{The electron-ion interaction leads to an induced attraction between electrons.
Arrows joining circles represent displaced positive ions that are attracted to the electron -
the timescale for relaxation back to their original positions is slow compared to the electron dynamics,
allowing a second electron to take advantage of this distortion.}
\label{e-p}
\end{figure}

The resulting strong coupling theory was developed by Schrieffer and colleagues \cite{SSW} into a precise formalism for
describing pairing in real systems.  The success of this theory was the prediction of anomalies in tunneling spectra
caused by the frequency dependence of the pairing self-energy associated with phonons that essentially proved that
conventional superconductivity originated from the electron-ion interaction \cite{rowell,mcmillan1}.  The theory also resulted
in a quantitative tool for estimating superconducting transition temperatures \cite{mcmillan2,dynes}.  From this, one can
understand what limits conventional superconductivity to relatively low temperatures \cite{cohen}.  In BCS theory, the
underlying mechanism is the electron-ion interaction.  An electron polarizes the surrounding lattice of ions.  Since the
ion timescale is much slower than the electrons (as they are much heavier), the polarization cloud persists as the
electron moves away.  A second electron can then move in and take advantage of this attractive polarization cloud (Fig.~\ref{e-p}).
This is how the electrons can indirectly attract each other despite the large Coulomb repulsion between them.  In essence,
the electrons avoid the Coulomb repulsion by being at the same place, but at different times.  There are two consequences
of this.  First, the electrons are in a s-wave pair state (which is a spin singlet due to fermion antisymmetry).  Second, 
the large Coulomb repulsion is renormalized to a smaller value when projecting from an energy scale $E_F$ down to a scale
$\hbar \omega_D$ \cite{morel,bauer}, thus allowing a net attraction, but the resulting `retardation' limits T$_c$.

But not all were so impressed by these developments.  The famous experimental physicist Bernd Matthias was well known 
for his negative opinion of 
BCS theory and its strong coupling avatars.  This came from a lack of prediction for any new superconductors.  The latter
was not a surprise given the exponential dependence of T$_c$ on microscopic parameters (a consequence of the logarithmic
infrared singularity), but Matthias' opinion was that if the strong coupling theory was so precise as claimed by its various
practitioners, why had it provided so little guidance to him and his experimental colleagues when searching for new superconductors?
In some sense, he went too far in asserting that only simple non-transition elements like mercury and lead were within
the sphere of BCS theory \cite{bernd1}.  It is now generally recognized that transition metals such as niobium and its higher
temperature A15 cousins like Nb$_3$Sn are well described by the Midgal-Eliashberg formalism \cite{doug}.  But the lack of
predictability is definitely an issue.  In that context, MgB$_2$ is a simple material that had been lying around for fifty years
before it was discovered to be a high temperature superconductor \cite{mgb2}.  Subsequently, it was shown that standard
strong coupling theory gave a good description of its properties \cite{mazin}.  But even predictions based on this success did
not pan out when looking for superconductivity in related materials \cite{pickett}.  This emphasizes that we have a long way to go 
before even conventional superconductivity becomes a truly predictive science.

So having emphasized `conventional', but do we mean by this and its counterpart `unconventional'?  In BCS theory, the pairing
is mediated by the electron-ion interaction, leading to a pair state with s-wave symmetry.  Anisotropy of the energy gap
(which is proportional to the superconducting order parameter in BCS theory) in momentum space is relatively weak.  But as soon
realized after the BCS theory was published, it could be easily generalized.  In BCS theory, the electron-ion interaction is
transformed into an effective electron-electron interaction limited to a shell in momentum space around the Fermi surface.
As such, any effective attractive interaction can be so treated, and it can even be extended to finite systems (such as the pairing
of nucleons in nuclei due to the strong interaction, where the `shell' in this case is the surface region of the nucleus \cite{BMP}).
Moreover, it can be easily generalized from an s-wave state to any other symmetry for the pair
state.  Therefore, by `unconventional', we mean a pair state that is not an isotropic s-wave state, and where the interaction is something
other than the conventional electron-ion interaction elucidated in the 1950s.

This brings us to $^3$He.

\section{Helium-3}

The first unconventional material didn't turn out to be a superconductor at all, but rather a superfluid.
As the BCS theory developed in the late 1950s and early 1960s, it was realized that it could be applied to a variety of interesting
systems.  It had already been known that $^4$He underwent Bose condensation at low temperatures.  But what about $^3$He?
As each atom is a fermion (two protons, a neutron, and two electrons), for it to condense, some kind of pairing must take place.
But how?  After all, these filled shell atoms have a large hard core repulsion.  But at larger separations,
an attractive van der Waals interaction exists.  By pairing in a d-wave state, the atoms could avoid the hard
core repulsion (since the d-wave state has a quadratic node at zero separation) and take advantage of the van der Waals tail 
(since the maximum of the d-wave state occurs in the tail region) \cite{emery}.

\begin{figure}
\includegraphics[width=1.5in]{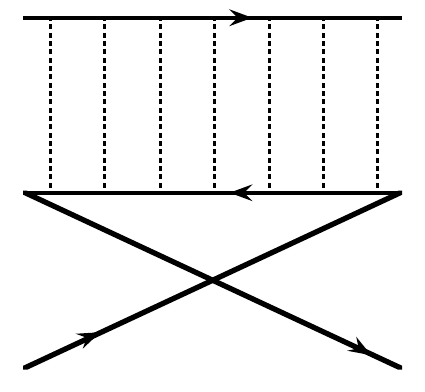}
\caption{Induced pair interaction from spin fluctuations \cite{doug12}.  Note the particle-hole ladder sum, which gives rise to the dynamic spin
susceptibility, embedded in this diagram.}
\label{ladder-sf}
\end{figure}

But in the late 1960s, a different potential mechanism was proposed.  To understand this one must go back to the early days of the BCS
theory.  Shortly after the BCS theory was published, Anderson realized that the state should survive even the presence of disorder,
since one can always define time reversed states even if the momentum states are smeared due to impurity scattering \cite{pwa59}.
Magnetic impurities, though, were different, in that they flipped the spin and thus broke the singlets \cite{AG}.  In strong coupling
theory, this pair breaking effect was easily generalized to inelastic ferromagnetic spin fluctuations \cite{BS}.  But Fay and Layzer \cite{fay}
realized that this argument could be turned around to argue that ferromagnetic spin fluctuations could mediate spin-triplet p-wave 
pairing (Fig.~\ref{ladder-sf}).  In essence, an `up' spin would prefer to have neighboring `up' spins, thus leading to an induced attraction due to exchange
forces.  They predicted that this could be the case for nearly ferromagnetic palladium (never realized, at least yet) as well as for $^3$He.

Regardless, the discovery of superfluidity in $^3$He in 1972 was a surprise \cite{ORL}.  The experimentalists were actually looking for magnetism
(which was subsequently found \cite{halperin}).  But what rapidly emerged was that they had indeed found p-wave superfluidity \cite{tony}.
And, it turned out that there were {\it two} superfluid phases.  The main phase was the so-called $B$ phase, first described theoretically
by Balian and Werthamer \cite{BW}.  In this phase, the pair state is of the form $k_x {\hat x} + k_y {\hat y} + k_z {\hat z}$ where ${\hat x}$,
for instance, means that the projection of the Cooper pair spin along this axis is zero (these three spin components form a vector known as
the $d$ vector).  Since the Fermi surface is a sphere, this function leads to an isotropic energy gap.  But, in a narrow sliver of 
temperature and pressure, another phase known as the $A$ phase exists.  This phase, first theoretically described by Anderson and Morel \cite{AM}
has the form  $(k_x + ik_y){\hat z}$.  This function has zeros (nodes) at the north and south `poles' of the Fermi surface, leading to a highly anisotropic
energy gap.

The existence of the $A$ phase was a surprise, since a simple Ginzburg-Landau (G-L) treatment would predict that the $B$ phase would always
be stable.  The reason is that its isotropic gap maximizes the free energy gain due to superfluidity (easily seen by evaluating
the quartic term in G-L theory).  A possible solution was offered by Anderson and Brinkman a year after the discovery of Osheroff,
Richardson and Lee \cite{pwa73}.  In spin fluctuation models, the pair interaction is strongly influenced by the superconductivity itself.
This is because the underlying fermion degrees of freedom become gapped, thus leading to a gap in the spin fluctuation spectrum,
which in turn suppresses the pairing.  This is very different from electron-ion theories, where the phonons do not become gapped (they do 
become less damped, of course, which does have a minor feedback effect on the pairing).  Obviously, this suppression effect is less
pronounced for the $A$ phase, given its anisotropic gap, which acts to stabilize the $A$ phase in a narrow temperature range (until
this feedback effect is overwhelmed by the quartic term which grows as the temperature is reduced).

The Anderson-Brinkman theory could have been viewed as such a success, one might simply have declared victory and moved on.
But life was not so simple for a variety of reasons.  $^3$He is a relatively simple system from a solid state physics perspective.  It is
a single band system with a simple parabolic dispersion, with weak spin-orbit effects.  The normal state interaction parameters (so-called
Landau parameters) are described by simple Legendre polynomials, and are well known from experiment.  These parameters in turn
determine the pairing interaction \cite{PZ}, which has been mapped to high precision \cite{serene}.  Analyzing in terms of physical
interactions, one finds that everything and the kitchen sink contributes to the pairing, including not only spin, but also density
and current fluctuations \cite{tony75}.

\begin{figure}
\includegraphics[width=1.5in]{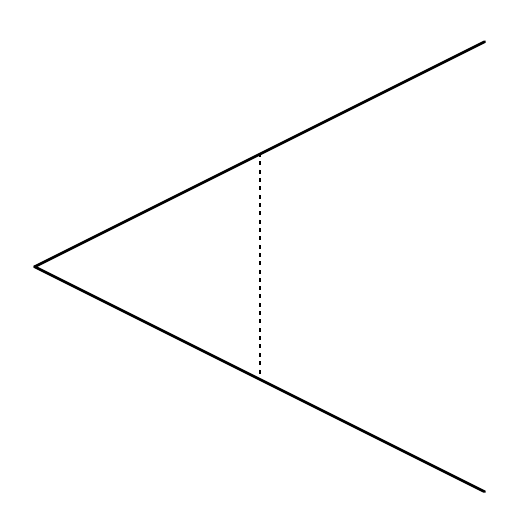}
\caption{Vertex correction \cite{vertex-L}.  In electron-electron theories, this diagram can be as large as the lowest
order self-energy (`rainbow') diagram, leading to a violation of Migdal's theorem.}
\label{vertex}
\end{figure}

This complexity has led to much richness in microscopic theories designed to explain $^3$He, which went on to play an important role
after the subsequent discovery of unconventional superconductors.  Strong coupling theories of spin fluctuations were developed to
further improve our understanding of $^3$He, and this led to one of the first occurrences of quantum criticality in the context of
superconductivity \cite{levin-valls}.  The idea was that as physical parameters such as pressure were tuned to approach the magnetically
ordered state, T$_c$ rose because of the increasing divergence of the pairing interaction, which in these models is proportional to
the dynamic spin susceptibility.  On the other hand, for the same reason, the energy scale of the spin fluctuations collapses as the critical point
is approached.  Eventually the latter effect wins out, and T$_c$, after achieving a maximum, is predicted to plummet to zero.  On the other hand, it
was also realized that the `Migdal-Eliashberg' basis of such calculations was suspect.  The reason is that the spin fluctuations are
composed of the same electrons that one is pairing, unlike the electron-ion case where electrons and phonons can be
considered as independent objects to a good precision.  One might naively think that one can separate the `slow' degrees of freedom
(the spin fluctuations) from the `fast' ones (quasiparticles), but explicit calculations find that vertex corrections can be of the same order
as the lowest order `rainbow' diagram (Fig.~\ref{vertex}) in stark contrast to the electron-ion case \cite{vertex-L}.  This casts doubts whether a controlled
perturbation expansion can be constructed, and the similar fears have recently been realized in the context of nematic and antiferromagnetic
spin fluctuations by Metlitski and Sachdev \cite{MS1,MS2}.  Besides spin fluctuations, other approaches have also been advocated, including
the polarization potential model of Bedell and Pines \cite{bedell}.

The other interesting point is that the magnetic phase of $^3$He is not a ferromagnet as centrally assumed in the spin fluctuation models.
It actually is an antiferromagnet \cite{roger}.  This led to a rethinking of the problem.  In particular, it was realized
that in some sense, the atoms in superfluid $^3$He are better described as nearly localized rather than nearly magnetic \cite{vollhardt}.
These ideas provided some of the foundational basis of theories that would later be developed in the context of cuprates - in particular
the idea of Gutzwiller projection (to suppress double occupation in the many body wavefunction) and the concept of using 1/$d$ as
an expansion parameter, where $d$ is the spatial dimension (the basis for dynamical mean field theory).

Most importantly, ideas in $^3$He paved the way for the next big development in unconventional superconductors - heavy fermions.

\section{Heavy Fermion Superconductors}

The discovery of heavy fermion superconductors again shows the hit and miss nature of the field of superconductivity.  Arguably, these
materials had been around for a number of years.  Despite the general prejudice that magnetism was bad for superconductivity \cite{AG,BS},
those like Matthias who were not enamored of the BCS theory thought otherwise.  Matthias had found superconductivity in a number of
uranium based intermetallics and pointed out the close connection of these materials with their ferromagnetic counterparts \cite{bernd-U}.
One of their more interesting discoveries was U$_2$PtC$_2$ \cite{bernd-U}.  But because elemental $\alpha$-U was suspected of being a conventional
superconductor (which it probably is), these results provoked less curiosity than they should have.  Then, in 1975, Bucher and colleagues
reported superconductivity in UBe$_{13}$, where the f electrons were nearly magnetic \cite{bucher}.  But their feeling was that the
superconductivity they observed was due to filaments of $\alpha$-U in their samples.

It took the remarkable discovery of superconductivity in CeCu$_2$Si$_2$ by Frank Steglich and collaborators to finally realize that a new
class of superconductors had been elucidated \cite{frank}.  By that time, it had been discovered that a number of rare earth and actinide intermetallics
exhibited a linear $T$ specific heat coefficient at low temperatures, typical of a Fermi liquid.  The difference, though, was that its magnitude was huge, up to
1000 times that of copper.  This indicated that the quasiparticles in such materials were strongly interacting, with the $f$ electrons being both
nearly localized and nearly magnetic.  In some sense, this would have been the last place one might expected to find a superconductor.
Moreover, Steglich's group realized that it was the heavy electrons themselves that were superconducting, since the jump of the specific heat at T$_c$ was
comparable to the normal state specific heat (this jump in BCS theory is proportional to the order parameter).

\begin{figure}
\includegraphics[width=3.4in]{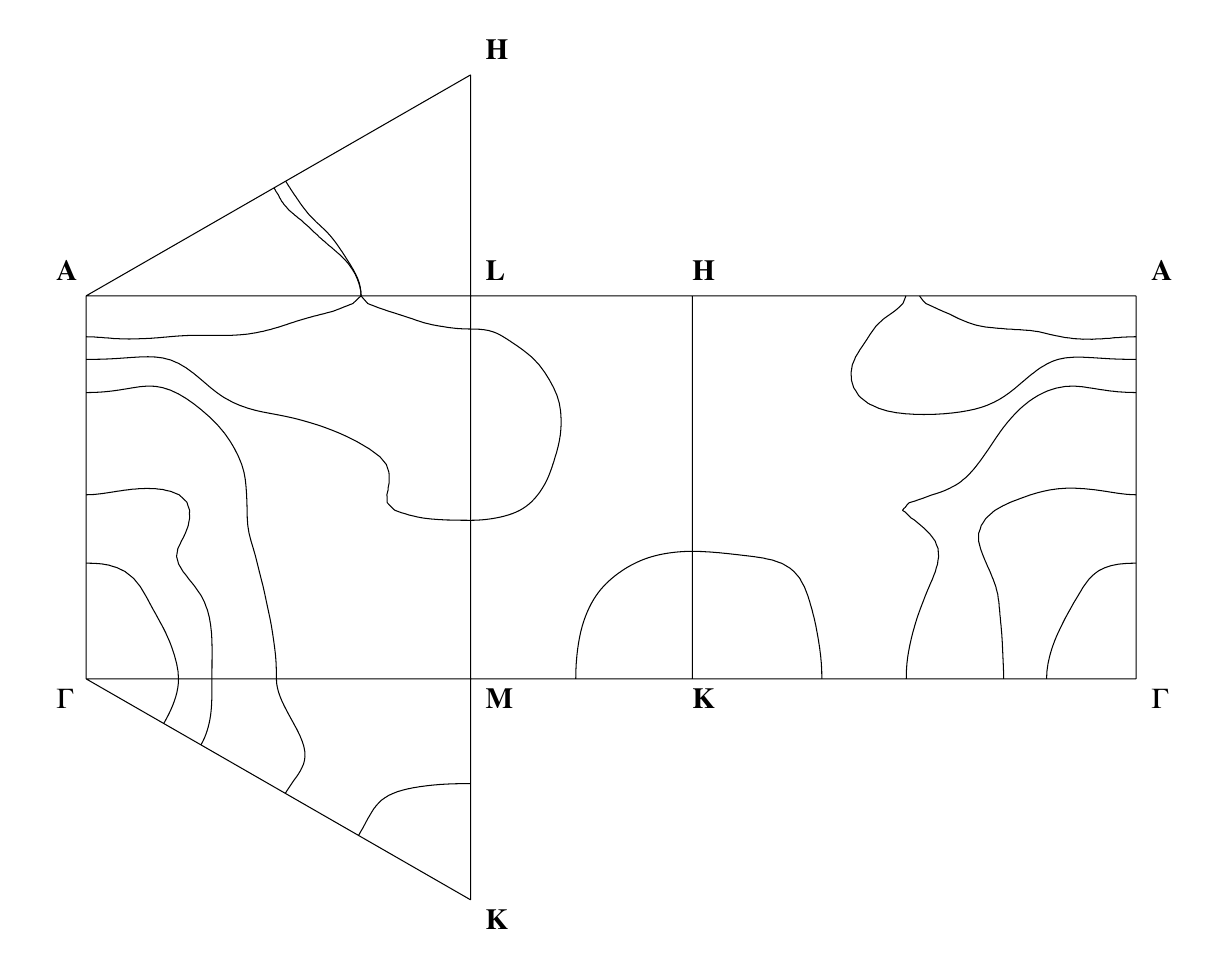}
\caption{Fermi surface of UPt$_3$ from local density band calculations \cite{albers}, plotted in the high symmetry
planes of the hexagonal Brillouin zone.  This is composed of four electron surfaces - three around $\Gamma$ and one around $K$ - and two hole
surfaces around $A$.}
\label{upt3-FS}
\end{figure}

After Steglich's breakthrough, much progress was made.  A few years later, the Los Alamos group discovered heavy fermion superconductivity in 
UPt$_3$ \cite{greg}, and
it was (re)discovered as well in UBe$_{13}$ \cite{ott}.  At this point, the field really took off.  Even U$_2$PtC$_2$ was realized to be
one as well \cite{gmeis}.  UPt$_3$ had properties reminiscent of $^3$He,
with what looked to be a T$^3$lnT correction to the specific heat (as predicted by spin fluctuation theories).  Moreover, the heavy quasiparticles
formed a normal (though complex!) Fermi surface (Fig.~\ref{upt3-FS}), as revealed by quantum oscillation measurements \cite{louis}.  This was a real 
surprise at the time,
since these measurements indicated that the $f$ electrons participated in the Fermi surface given the latter's resemblance to simple band theory
calculations which treated the $f$ electrons as itinerant \cite{albers}.  Although over the years, alternate models of the Fermi surface were proposed,
with some $f$ electrons participating in the Fermi surface and others not \cite{gertrud}, recent definitive results decisively verify the itineracy of all the
$f$ electrons \cite{McM}.

\begin{figure}
\includegraphics[width=3.4in]{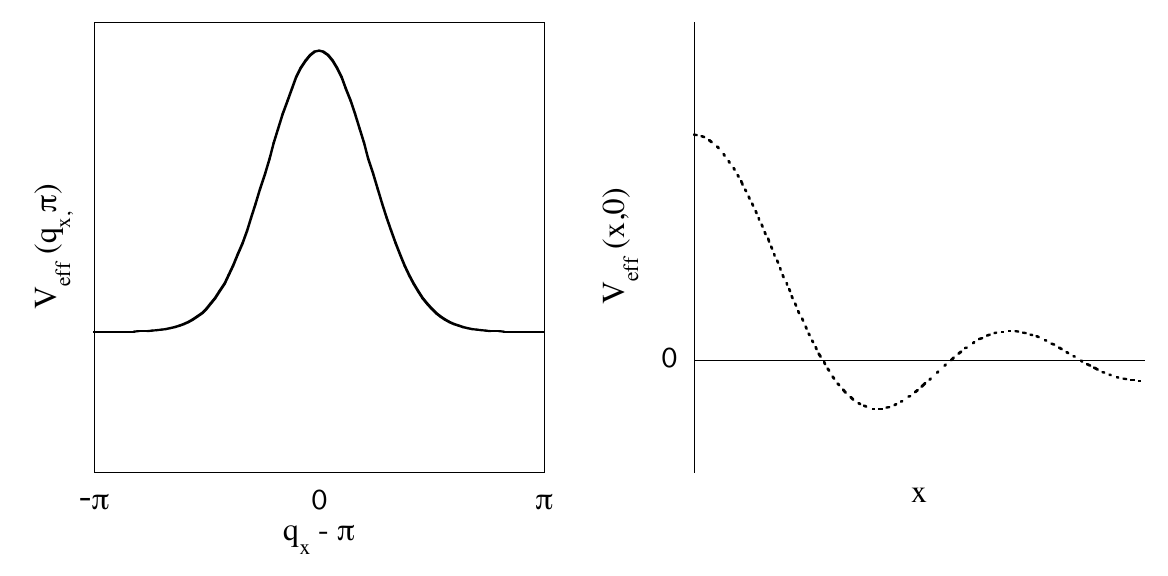}
\caption{Induced pair interaction from antiferromagnetic spin fluctuations \cite{doug12}.  Momentum space (left)
with a repulsive potential peaked at the magnetic wavevector $Q$.  Fourier transform (right) showing Friedel-like
oscillations, with a repulsive on-site potential, and an attractive potential for near-neighbor separations.}
\label{SF-pairing}
\end{figure}

Given the resemblance to $^3$He, it was not surprising that theorists tried to translate theories for $^3$He over to the heavy fermion case.  But
trouble soon brewed.  Neutron scattering measurements indicated the presence of antiferromagnetic spin fluctuations, {\it not} ferromagnetic
spin fluctuations, in UPt$_3$ \cite{aeppli}.  Several groups then realized that this difference would lead to d-wave singlet pairing instead of p-wave
triplet pairing \cite{SLH,MSV,BBE,SLH2} (Fig.~\ref{SF-pairing}).  In a single band simple cubic model, the pair state would be of the 
form $d_{x^2-y^2} \pm i d_{3z^2-r^2}$.
In real space, this corresponds to lobes that point from one atom to its six surrounding neighbors.

But there was the rub.  UPt$_3$, even in a band theory description, is a very complex beast.  Six j=5/2 $f$ bands are in the vicinity of the Fermi
energy (spin-orbit coupling being particularly strong).  Of these, five are predicted to cross the Fermi energy.  These bands are complicated
admixtures of these $f$ orbitals with uranium 6d and platinum 5d orbitals.  Even constructing a pairing interaction at the phenomenological level
is difficult, as two types of antiferromagnetic fluctuations are seen.  The original `high energy' ones correspond to antiferromagnetic correlations
between near neighbor uranium ions \cite{aeppli}.  But after this, lower energy fluctuations were seen corresponding to antiferromagnetic correlations between
next near neighbors \cite{broholm}.  This frustrated interaction leads to stripe-like order, lowering the symmetry from hexagonal to orthorhombic.
UPt$_3$ actually quasi-orders at this wavevector well above T$_c$, but with a tiny moment (that becomes large only when doped with other ions,
like palladium).  True long range order only sets in well below T$_c$ \cite{schubert}.

Given these complications, what was done was to construct a model for the pair state based on experimental data.  This led to surprising directions.
First, UPt$_3$ does not exhibit any change in the Knight shift when going below T$_c$ \cite{tou}.  This implies that the pair state is a triplet, certainly
not what would naively expect based on antiferromagnetic fluctuations.  Also, various measurements, such as thermal conductivity and transverse
ultrasound, are most consistent with the presence of nodes (where the order parameter vanishes) on lines on the equator of the Fermi surface, along with
`quadratic' point nodes at the north and south poles where the gap varies quadratically with the polar angle \cite{sauls,mike92}.
In terms of spherical harmonics,
the first one encountered with this property is Y$_{32}$.  This is from the E$_{2u}$ representation of hexagonal symmetry.  It forms a `triplet' state when
multiplying this by ${\hat z}$.  This acts to project the Cooper pair spins into the basal plane, consistent with the normal state spin susceptibility,
and also the upper critical field, which indicates Pauli limiting (Zeeman pair breaking) for fields along the $c$ axis (for in-plane fields, the Cooper pair 
spins can obviously align with the field direction).  Note this `triplet' is actually a `singlet' when counting spin degrees of freedom.  It is thought
that the `locking' of the $d$ vector to the c axis is a consequence of the strong spin-orbit coupling in UPt$_3$.

\begin{figure}
\includegraphics[width=1.5in]{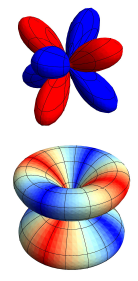}
\caption{E$_{2u}$ (f-wave) order parameter \cite{strand1}.  The bottom plot is Y$_{32}$, that is $k_z(k_x + i k_y)^2$.
The top one is the real part of this.}
\label{e2u}
\end{figure}

\begin{figure}
\includegraphics[width=3in]{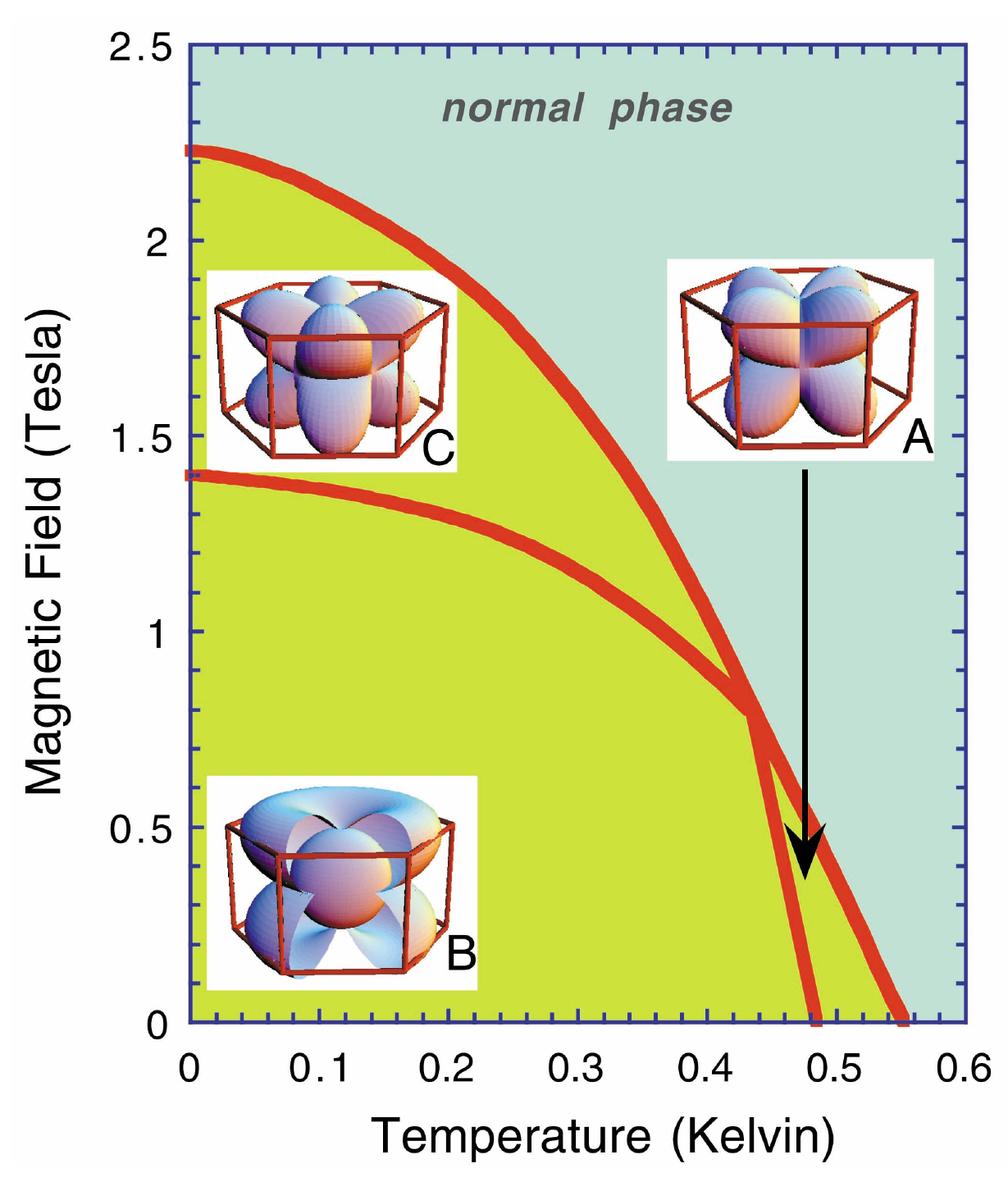}
\caption{Phase diagram of UPt$_3$ versus field \cite{huxley}, exhibiting three superconducting phases - A, B and C.
In the E$_{2u}$ model, the A phase would correspond to the top plot in Fig.~\ref{e2u},
the B phase to the bottom one.}
\label{upt3}
\end{figure}

There are a number of other advantages of this E$_{2u}$ model  (Fig.~\ref{e2u}).  Several years after the discovery of superconductivity in UPt$_3$,
{\it two} superconducting phase transitions were discovered \cite{split-phase}, again very reminiscent of $^3$He.  Then, it was realized that in a magnetic field, yet
another transition takes place \cite{hassel,aden,huxley}, making three superconducting phases altogether.  Note that this last phase transition occurs for fields between 
the lower and upper critical fields associated with the vortex phase, and is thought to represent another distinct phase in the relative degrees of
freedom (as opposed to a change in the vortex lattice, that would be associated with the center of mass degrees of freedom of the pair state).
The resulting phase diagram in the $H$-$T$ plane (Fig.~\ref{upt3}) exhibits an unusual `tetracritical' point where the three superconducting phases 
and the normal phase touch.

There are two basic models that can explain these observations.  First, two nearly degenerate solutions, ironically known as the $A$-$B$ 
model \cite{garg} though for different reasons than $^3$He.  Here, $A$ refers to one of the $A$ single dimensional
representations of the hexagonal group, and $B$ to one of the $B$ single dimensional representations.  An advantage of this model is that it allows
the tetracritical point to exist, since the $A$ and $B$ phase lines can cross since they come from different group representations, though the near
degeneracy of the two solutions is an assumption with no real microscopic basis.  The alternate
model is for the order parameter to come from a two dimensional group representation, like E$_{2u}$.  The advantage of this model is that it
naturally explains the near degeneracy of the two zero field phase transitions.  A likely source for the small degeneracy lifting is the small moment
magnetism mentioned above, which leads to weak orthorhombicity.  In support of this, under pressure, the double transition goes away at essentially
the same pressure that the magnetism disappears \cite{hayden}.  On the other hand, it is non-trivial for this model to account for the tetracritical point,
since the various phases all originate from the same group representation, leading to level repulsion and thus an avoided crossing rather than
a point of degeneracy.  But this is a potential advantage of the E$_{2u}$ model, in comparison to related models based on the d-wave 
E$_{1g}$ \cite{joynt} or p-wave E$_{1u}$ \cite{machida} models.  For the latter, the two bases of the two dimensional representation differ by two 
units of angular momentum (that is Y$_{2\pm1}$ for E$_{1g}$, and Y$_{1\pm1}$ or Y$_{3\pm1}$ for E$_{1u}$).
The result is that gradient terms in the G-L free
energy couple the two bases, leading to a splitting of the tetracritical point.  On the other hand, the two partners in the E$_{2u}$ case
differ by four units of angular momentum (Y$_{3\pm2}$), and thus in an axial approximation, no splitting occurs.  Hexagonal anisotropy will couple
the two, but the hope is that this is weak, which is supported by explicit calculations \cite{VV}.
Strong support for the E$_{2u}$ model has recently come from phase sensitive Josephson tunneling, which is consistent with a two dimensional
representation with bases each having two units of angular momentum \cite{strand1,strand2}.  These measurements are also consistent with the predicted
nodal structure of this state.

\begin{figure}
\includegraphics[width=2in]{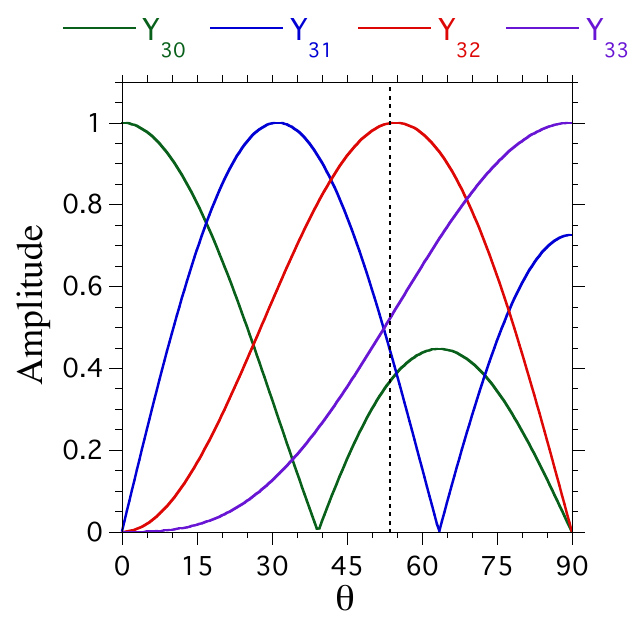}
\caption{Amplitude of Y$_{3m}$ versus polar angle.  The vertical dashed line is the angle corresponding
to that of near neighbor uranium atoms in UPt$_3$, which is close to the maximum of Y$_{32}$.}
\label{ylm}
\end{figure}

Still, a number of important questions remain.  Both the phase sensitive tunneling and transverse ultrasound \cite{TU} indicate a single
domain state, whereas any of the above two dimensional models would predict three different hexagonal domains.
Why only one domain is realized in a macroscopic
sample, and at that the `right' one (the predicted transverse ultrasound only agrees with experiment for one of the the three domains \cite{sauls-US})
is unknown.
The $d$ vector structure has also been brought into question \cite{machida}, since the Knight shift indicates no change below T$_c$ for any 
field orientation,
which is most easily explained if the $d$ vector can be rotated by the field.  Even the nodal structure for E$_{2u}$ has been recently 
questioned \cite{louis-node,tsutsumi}.
Still, if it is an f-wave E$_{2u}$ state, there may be a relatively simple explanation for it.  Plots of the spherical harmonics versus polar angle find
that the maximum of the Y$_{32}$ harmonic is close to the angle separating near neighbor uranium atoms in UPt$_3$ (Fig.~\ref{ylm}).
Therefore, if such a state is realized, it is in strong support of pairing models based on near neighbor interactions, such as occurs in models
based on antiferromagnetic spin fluctuations.  But then why a triplet in that case?

To understand this, we need to step back and look at the general problem of pairing in the presence of strong spin-orbit coupling, where spin is no
longer a good quantum number.  This was first addressed by Anderson \cite{pwa84}.  What he realized was that for a given state $k$, one can still
define four degenerate states (assuming time reversal and inversion symmetries are not broken): $k, Pk, Tk, PTk$ where $P$ is the parity operator
and $T$ the time reversal one.  From these four states (corresponding in the spin-only case to $k\uparrow, -k\uparrow, -k\downarrow, k\downarrow$)
one can construct a `singlet' and a `triplet'.  These states are respectively even parity and odd parity due to fermion anitsymmetry.
This formalism has been exploited in great detail by a number of authors  to understand the general structure
of the order parameter \cite{sigrist}.  What one finds is that the same spin-orbit effects that were invoked above to lock the $d$ vector to the lattice
also act to mix in the other two $d$ vector components (since spin is not a good quantum number).  The requirement that all three components 
vanish can only occur on points on the Fermi
surface, which is known as Blount's theorem \cite{blount}.  This would seem to eliminate any odd parity state description for UPt$_3$ if line nodes
are indeed present, except there
is an exception to the theorem.  Pair states are classified by representations at the $\Gamma$ point of the Brillouin zone since the
center of mass momentum of the pairs is zero (though finite momentum pair states for UPt$_3$ have been advocated \cite{cox}).  What this does
not take into account, though, is the composite nature of such pairs.  For non-symmorphic space groups (those with screw axes or glide planes), $k$
states on the zone boundary have special properties.  UPt$_3$ is an example, being a hexagonal close packed lattice with a screw axis.
This leads to bands which stick in pairs on the zone face perpendicular to the c axis \cite{herring} (this phenomenon is responsible for the
magnetic breakdown orbits seen in quantum oscillation measurements \cite{louis,albers}).  The same phenomenon causes all three $d$ vector 
components to vanish on the zone face for certain odd parity representations \cite{norm95,micklitz}.  Therefore, those Fermi surfaces which cross
the zone boundary can indeed have line nodes for a general pair state involving all three $d$ vector components.  Interestingly, these general pair states
bear little resemblance to the spherical harmonics mentioned above \cite{norm94,peter96}, and therefore one should view with caution statements
that material $X$ has `p', `d', or `f' wave pairing.

Much time has been spent discussing the UPt$_3$ case since it is an illustrative example of what is involved when discussing a complex multi-band
material in the presence of strong spin-orbit coupling.  But there are many other examples of heavy fermion superconductors which reveal a
great wealth of phenomena.  UBe$_{13}$ was discovered at about the same time as UPt$_3$, but it seems to be a very different animal.  Unlike
UPt$_3$, where Fermi liquid like behavior sets in well above T$_c$, in UBe$_{13}$ it never sets in \cite{ott}.  That is, the superconductor emerges from a 
non-Fermi liquid normal state.  This is profound, since superconductivity is an instability of the normal state, and the underlying supposition of BCS 
is that the normal state is composed of quasiparticles.  Little is known about the superconducting state, though it appears that again, there are multiple
superconducting phases (this particularly becomes evident when one dopes with thorium \cite{th-ube13}).

At about the same time as the materials discussed above, superconductivity was discovered in 
URu$_2$Si$_2$ \cite{maple}.  This material continues to be fascinating because of the
unknown nature of its normal state.  At 17K, a transition occurs to what was thought at the time to be an antiferromagnetic state.  Yet subsequent
neutron scattering measurements found the ordered moment to be tiny \cite{urs-ns}, far too small to explain the large specific heat anomaly that indicates
that most of the Fermi surface has been removed.  After many years, it was realized that internal strain was responsible for the small moments, 
and therefore the `hidden order' phase is not magnetic, though it appears to be related to an antiferromagnetic phase, which can be induced
by either doping or pressure.  What it is still remains a point of great speculation \cite{mydosh}, and until this is resolved, the nature of the superconducting
state will be difficult to resolve as well.  One of the most intriguing suggestions is that the hidden order is due to some higher multipolar 
order \cite{haule,lars,ikeda}, though to date, no evidence of this has been forthcoming from x-ray measurements.

There are, though, close cousins which exhibit robust antiferromagnetic order:  UPd$_2$Al$_3$ and UNi$_2$Al$_3$ \cite{updal}.  That is, the
superconducting state emerges from an antiferromagnet.  Knight shift measurements are consistent with a `singlet' for the former but a `triplet' for
the latter \cite{updal-K}.  Although the presence of antiferromagnetism does break time reversal symmetry, singlet pairing is still possible - the
actual order parameter being a linear combination of a spin singlet with zero center of mass momentum and one component of a spin triplet with a
momentum equal to the antiferromagnetic wavevector \cite{fenton}.  UPd$_2$Al$_3$ was the first heavy fermion superconductor to exhibit a
spin `resonance' as seen by inelastic neutron scattering.
$^3$He has many collective modes of the superconducting order parameter \cite{V-W}, but this is due to strong degeneracy of
the order parameter (three orbital times three spin degrees of freedom) and its neutral nature.  In conventional charged superconductors, 
collective modes have not been found, except for the Carlson-Goldman mode \cite{CG} (a `phase' mode which occurs near T$_c$ because
of backflow of the normal carriers \cite{schon}), the Higgs mode \cite{schmit} (which only becomes a true collective mode when it is pulled
below the 2$\Delta$ continuum due to interactions, as occurs when superconducting and charge density wave order coexist \cite{klein,LV}), and
the `Leggett' mode \cite{tony66} (where the relative phase of the order parameter of a multi-band system can oscillate, as thought to have been
seen by Raman scattering in MgB$_2$ \cite{girsh}).

The suppression of collective modes can be understood from the BCS coherence factors. The polarization bubble in the superconducting phase 
is composed of two terms:  $GG$ and $FF$, where $G$ is the normal and $F$ the anomalous (Gor'kov)
Green's function.  Typically, these two contributions cancel on the mass shell, leading for instance in the s-wave case to a square root growth in frequency
of the conductivity above the 2$\Delta$ threshold (the `missing' weight shows up as the condensate peak at zero frequency).  But if the order
parameter should change sign under translation by a given $Q$ vector, then in the resulting finite momentum response, $GG$ and $FF$ reinforce one another rather
than cancel since the sign of $FF$ flips.  This leads to a step jump in the imaginary part of the bubble at 
$\hbar \omega_{th} = {\rm min}_k(|\Delta_k| + |\Delta_{k+Q}|)$, which by the Kramers-Kronig relation leads to a log divergence in the real part.
This divergence causes a pole to be pulled below the continuum within a linear response (RPA) treatment where $\chi = \chi_0/(1-I\chi_0)$
with $\chi_0$ the bubble and $I$ the exchange interaction in the case of the dynamic spin susceptibility.
Thus, the observation of a spin resonance in UPd$_2$Al$_3$ at the antiferromagnetic wavevector implies
the existence of such an order parameter \cite{nick}, though alternate possibilities have been suggested (magnons associated with
the ordered magnetic phase become less damped below 2$\Delta$ \cite{thalmeir}).

\begin{figure}
\includegraphics[width=2.75in]{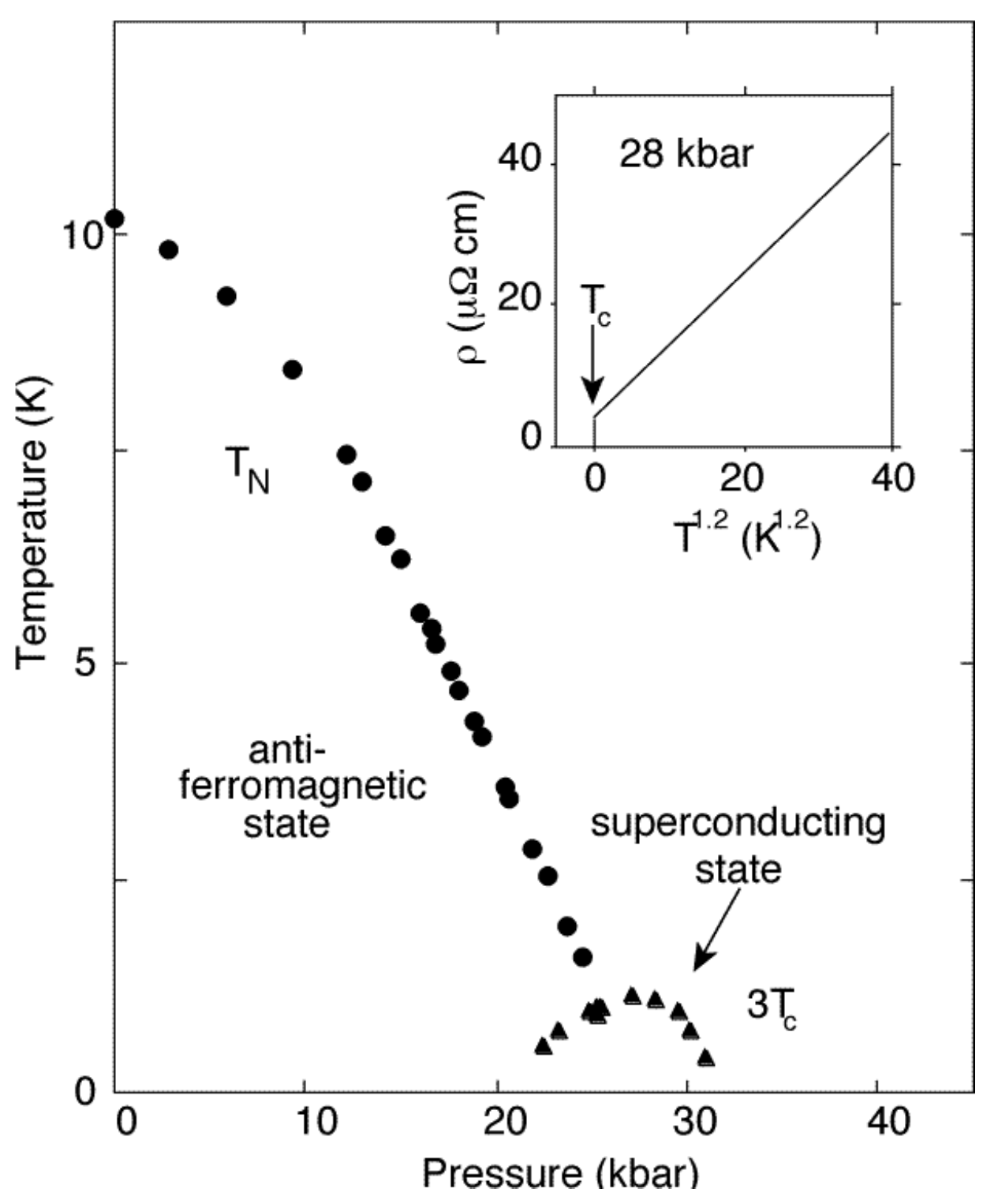}
\caption{Phase diagram of CePd$_2$Si$_2$ versus pressure \cite{mathur}.  The inset shows the resistivity at 28 kbar, which is quasi-linear in $T$.}
\label{cein3}
\end{figure}

The question of whether magnetic correlations are responsible for heavy fermion superconductivity provided a guiding principle when looking
for new ones.  In a classic paper \cite{mathur}, Gil Lonzarich's group demonstrated that the antiferromagnetic phases in CeIn$_3$ and CePd$_2$Si$_2$
were suppressed with pressure.  At the `quantum critical' point where the order was suppressed to zero, a `dome' of superconductivity appeared
(Fig.~\ref{cein3}).  This implied that quantum critical fluctuations associated with the magnetic order might potentially be the pairing `glue'.
Interestingly, this physics is similar to that proposed by Levin and Valls for $^3$He \cite{levin-valls}, yet in this case, the maximum $T_c$ appeared to
be {\it at} the critical point rather than displaced to the paramagnetic side as they predicted.

\begin{figure}
\includegraphics[width=2.75in]{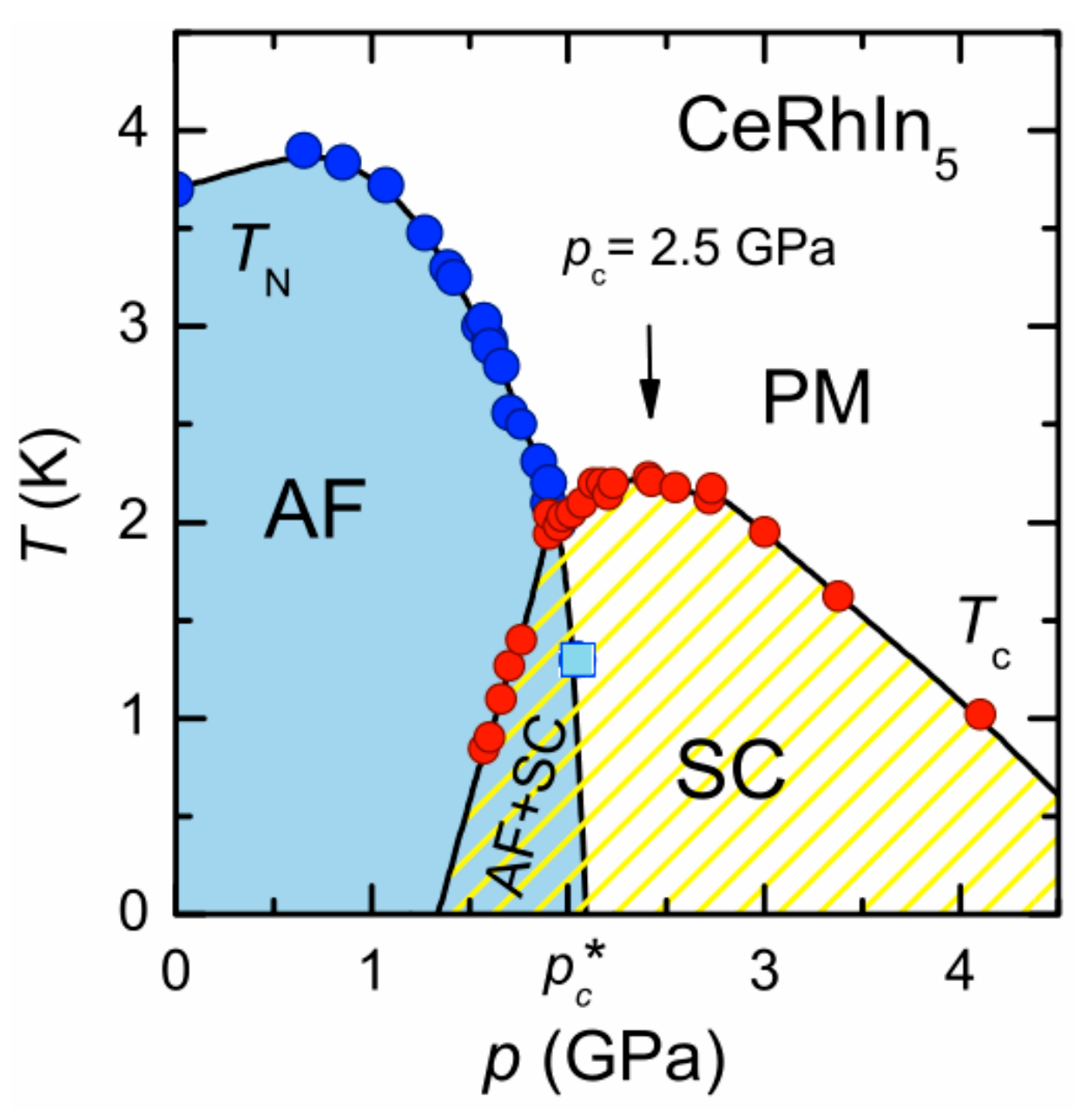}
\caption{Phase diagram of CeRhIn$_5$ versus pressure \cite{knebel}.  Note the coexistence region of antiferromagnetism and superconductivity.}
\label{phase-115}
\end{figure}

The discovery of Mathur \etal~promoted a resurgence in the field of heavy fermion superconductivity.  A few years later, superconductivity was
discovered in the CeXIn$_5$ compounds \cite{115}, where $X$ is a transition metal (Co, Ir, Rh). These materials are layered analogues of
cubic CeIn$_3$, and show superconducting phases overlapping with antiferromagnetic phases (Fig.~\ref{phase-115}).  Recently, the
superconductivity was seen to persist to just a few layers \cite{matsuda}.  Perhaps more dramatically, a plutonium analogue was found
to superconduct at 18.5 K \cite{sarrao}.  This $T_c$ was almost an order of magnitude larger than any previously known heavy fermion
superconductor.  NMR measurements for all of these materials indicate `singlet' pairing, and evidence for order parameter nodes have been provided 
by a variety of  measurements.  Therefore, in the literature, these have been referred to as `d' wave superconductors, with the caveat that the notation 
may be somewhat misleading because of the multi-band nature of these materials along with strong spin-orbit coupling.  One of the more
intriguing aspects is a new phase that appears at low temperatures just below the upper critical field in CeCoIn$_5$ \cite{radovan}.
In the beginning, it was felt this might be the long predicted `LOFF' state (where the electrons pair at finite momentum to help offset the
deleterious effects of the Zeeman splitting on the pairs), but recent neutron data point instead to a novel magnetic phase that is only stable below
the upper critical field \cite{kenz}.

\begin{figure}
\includegraphics[width=3in]{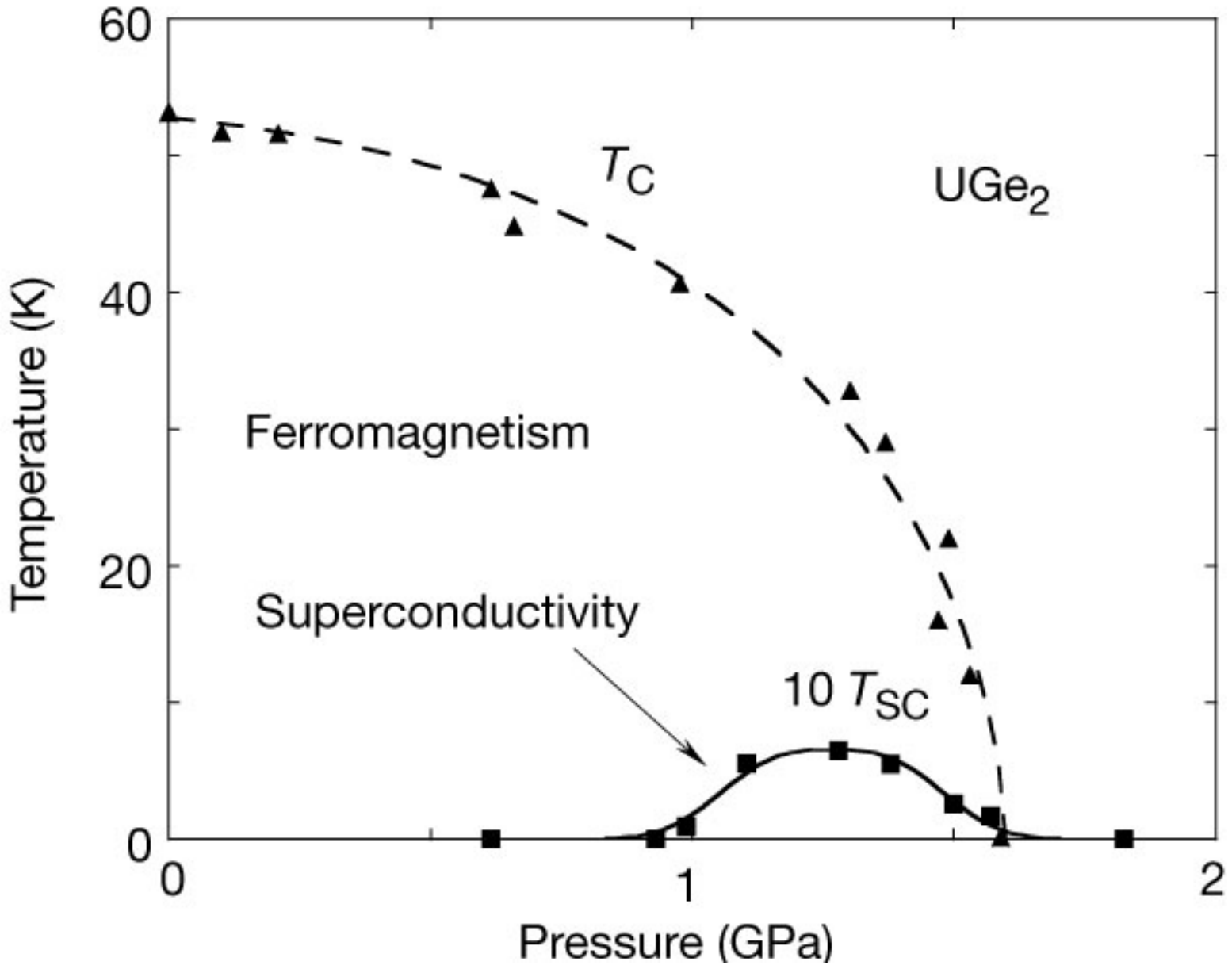}
\caption{Phase diagram of UGe$_2$ versus pressure \cite{uge2}.  The superconducting dome is completely inside the ferromagnetic phase.}
\label{uge}
\end{figure}

The other unusual discovery was that of superconductivity in UGe$_2$ \cite{uge2} and URhGe \cite{urg}.  These materials are {\it  ferromagnetic}.
Moreover, the superconducting `dome' (T$_c$ versus pressure) in UGe$_2$ is enclosed entirely within the ferromagnetic phase (Fig.~\ref{uge}).
URhGe exhibits an unusual `reentrant' behavior where upon applying a magnetic field, superconductivity is suppressed, then reappears at a higher
field \cite{urh-H}.  Obviously, the pair state in these materials is thought to be a `triplet', but little is known about its properties \cite{ferro-rev}.

We now turn to microscopics.  Heavy fermion behavior typically occurs near the borderline between localized and itinerant behavior for the $f$
electrons, the so-called Hill limit \cite{hill}.  Rare earth impurities in transition metals are well known to exhibit the Kondo effect, where scattering
of the conduction electrons off the $f$ ions leads to a logarithmic divergence of the resistivity as the temperature is lowered \cite{kondo}.
This is a good example of where perturbation theory
breaks down in an unusual way (it is third order in the interaction before the log shows up).  The Kondo problem was first solved by Ken Wilson
in 1975 using the numerical renormalization group \cite{KW}, where below the so-called Kondo temperature, the conduction
electrons bind to the $f$ electrons to form singlets, reminiscent of BCS theory.  This can presumably be extended to a dense array of such 
local ions, forming a `Kondo
lattice'.  Realistic treatments of the problem are based on the Anderson model \cite{pwa}, which
allows the $f$ occupation to be non-integer and thus accounts for $f$ charge fluctuations - the Kondo limit being the limit that the $f$ occupation 
goes to an integer value (i.e., the Coulomb repulsion $U$ goes to infinity), and thus only $f$ spin fluctuations remain.
The solution of this problem can be seen as a local $f$ level which interacts
with the conduction band, forming two `hybridized' bands (this is a correlated analogue of band theory).  If the chemical potential falls inside
the gap (integer occupation of $f$ and conduction), one has a `Kondo insulator' (currently the rage because it is a potential topological 
insulator with conducting surface states \cite{dzero,wolgast}).  If just outside the gap, one has a very heavy mass.

One can go beyond this mean
field treatment by the use of slave bosons with a gauge field that incorporates the constraint of near integer occupation of the $f$ electron (with the
scalar part of the gauge field related to the $f$ charge, and the vector part related to the $f$ current) \cite{slave}.  In this case, a perturbation
expansion is possible in 1/$N$, where $N$ is the degeneracy of the $f$ orbitals.  $N$ is six for the j=5/2 orbitals appropriate for cerium, 
but obviously in the low energy limit, $N$ typically reduces to 2 because of crystal field splitting of the $f$ levels.  The principal fluctuations beyond mean
field theory are hybridization fluctuations.  By considering the anomalous self-energy, these `Kondo bosons'
can intermediate higher angular momentum pairing \cite{lavagna}.  One disadvantage of this approach is that
spin fluctuations do not show up until order $1/N^2$ \cite{houghton}.  This can be cured by going to a spin rotationally invariant formalism.

\begin{figure}
\includegraphics[width=3.2in]{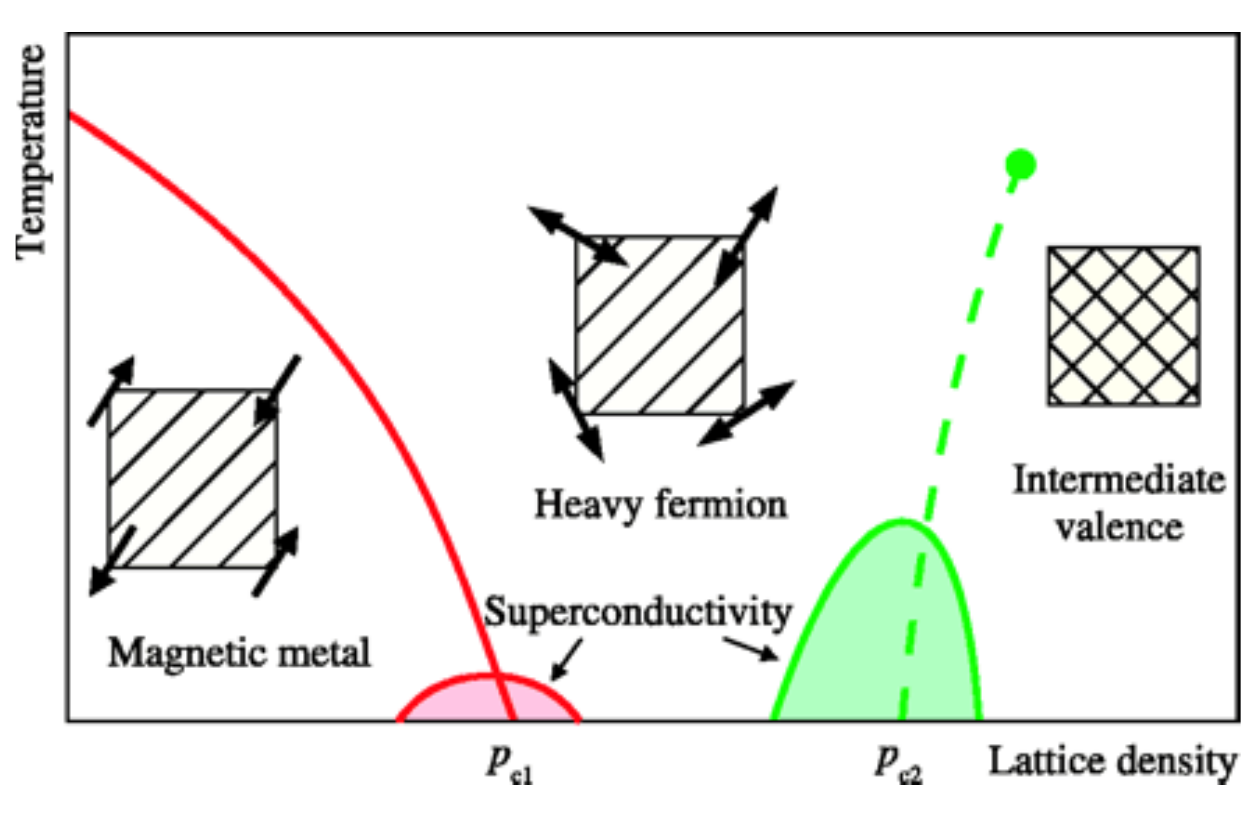}
\caption{Schematic phase diagram of CeCu$_2$Si$_{2-x}$Ge$_x$ versus pressure \cite{yuan}.
Two superconducting domes are present, the left one associated
with a quantum critical magnetic point, the second with a volume
collapse transition.}
\label{ccs}
\end{figure}

Since these early days, many theories for heavy fermion superconductivity have been proposed, ranging from the paramagnon and `Kondo boson'
approaches mentioned above, to phonons and valence fluctuations.  That phonons could play some role is evident from the very large Gruneisen
parameters observed in heavy fermion metals.  That valence fluctuations can play some role is evident from the phase diagram of
CeCu$_2$Si$_2$ (Fig.~\ref{ccs}).  The pressure dependence of T$_c$ is complicated, but upon doping with germanium (which
suppresses T$_c$), it was seen that the superconducting `dome' was actually composed of two domes \cite{yuan}.  The first (smaller) one is associated
with a quantum critical point where magnetic order disappears similar to CeIn$_3$, but the second (larger) one appears to be associated with
a valence change of the $f$ electrons.

\begin{figure}
\includegraphics[width=3.4in]{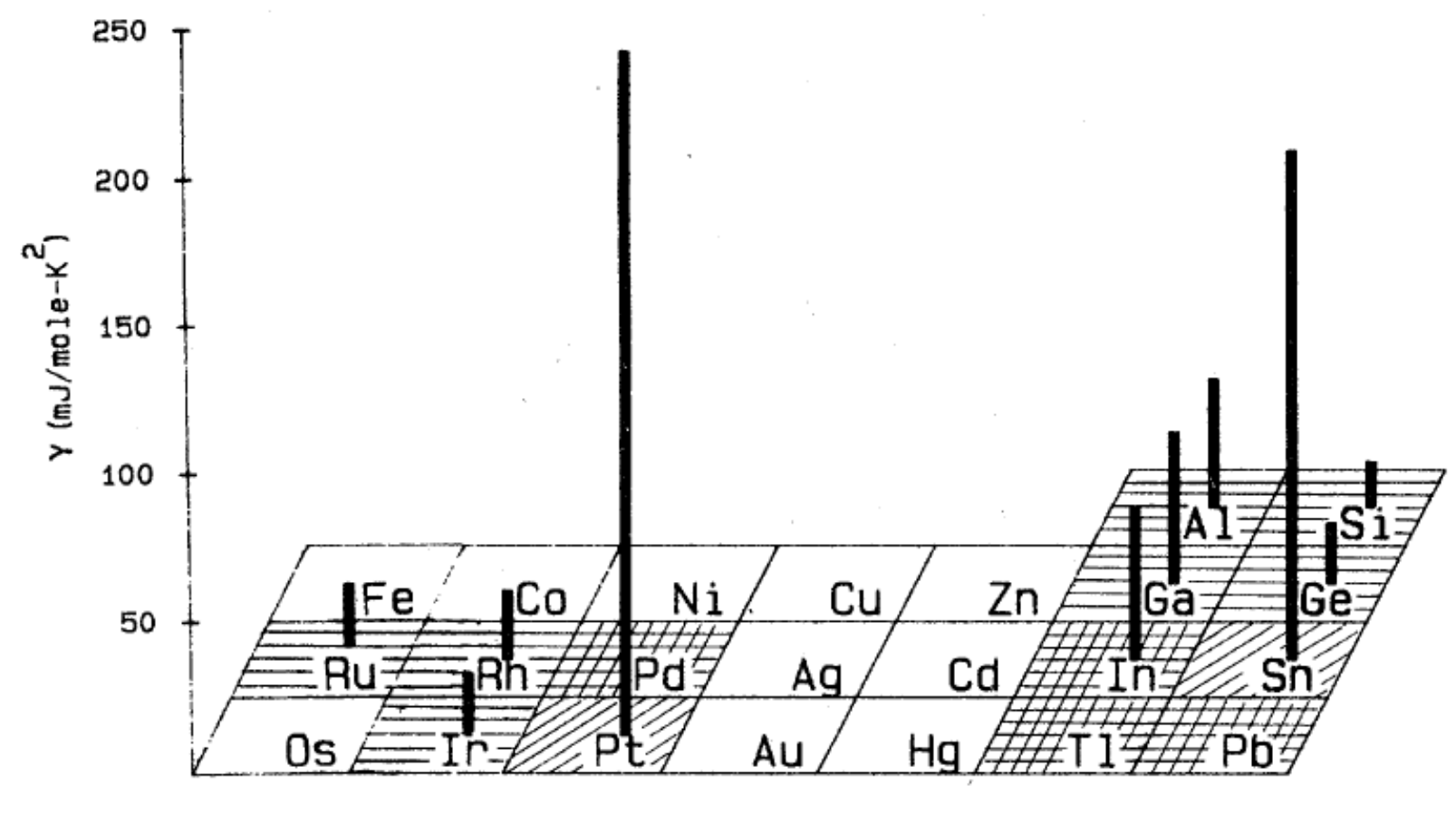}
\caption{Low temperature specific heat coefficient for various UX$_3$
alloys \cite{dale}.  All have the cubic AuCu$_3$ structure except for quadrupolar
ordered UPd$_3$ (double hexagonal close packed) and superconducting
UPt$_3$ (hexagonal close packed).}
\label{ux3}
\end{figure}

Reviewing the full breadth of these theories would take its own review article.  Suffice it to say that as of yet, there is no predictive theory that has 
emerged - some heavy fermion systems are magnetic (some of those even exhibiting itinerant spin density wave behavior), some are superconducting, 
and some are `vegetables', and we have little feeling for why this is so.  What is known is that heavy fermion superconductivity seems to occur under 
special conditions.  For instance, Fig.~\ref{ux3} shows the UX$_3$ materials \cite{dale}.  Almost all of them have the cubic AuCu$_3$ structure, and exhibit a wide 
range of behavior, but none of them are superconducting.  The exception structure wise is UPd$_3$ (double hexagonal close packed) which exhibits 
quadupolar order of localized  $f$ electrons, the other is superconducting UPt$_3$ (hexagonal close packed).  Why certain crystal structures seem to 
be amendable to superconductivity is not known - for instance, CeCu$_2$Si$_2$ and URu$_2$Si$_2$ both have the ThCr$_2$Si$_2$ structure,
as does BaFe$_2$As$_2$ (which becomes a high temperature superconductor upon doping as will be discussed below).

\section{Cuprates}

\begin{figure}
\includegraphics[width=2.25in]{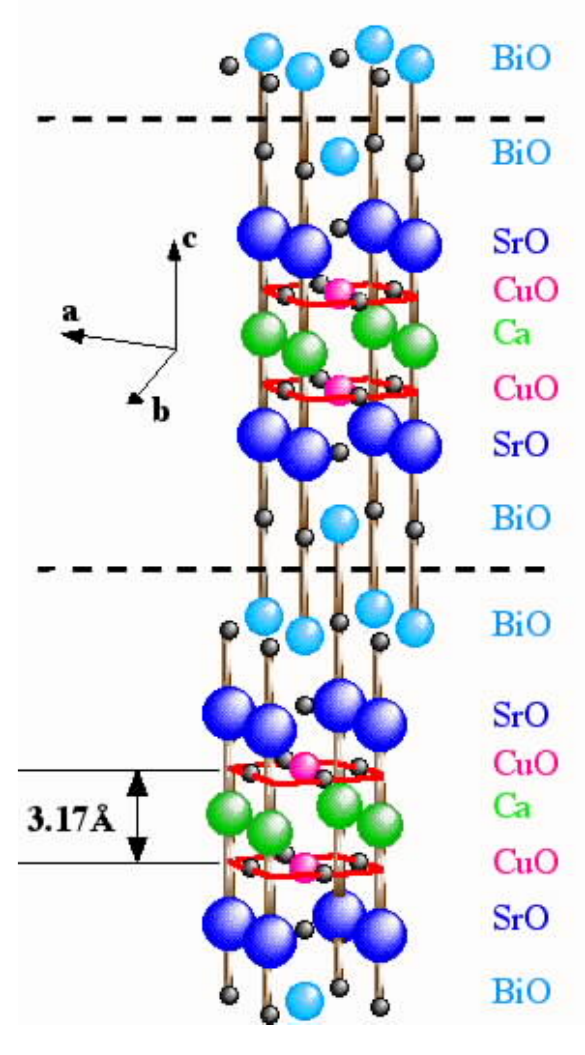}
\caption{Crystal structure of the cuprate Bi2212.  Bilayers of CuO$_2$ units are separated by
insulating spacer layers of SrO and BiO.  The dashed lines indicate a well defined cleavage plane, making
this material ideal for ARPES and STM studies.  Figure courtesy of Adam Kaminski.}
\label{cuprate-CS}
\end{figure}

V$_3$Si was discovered in 1953 \cite{hardy}, and this class of A15 cubic materials had the highest known T$_c$ (23 K for Nb$_3$Ge) until 
the cuprates were discovered in 1986 \cite{bednorz}.  This long stretch of time was what led to theoretical speculations that this might be the highest one 
might ever get to \cite{cohen}.  In fact, in the beginning, few people paid attention to the Bednorz-Muller paper on Ba doped La$_2$CuO$_4$ since
in the past, there had been so many sightings of `superconductors'
which had turned out to be false (so-called USOs - unidentified superconducting objects).  But about six months after their discovery,
their finding was verified by Tanaka's group, and progress became rapid, with the identification of superconductivity above liquid nitrogen temperature
a few months later in the related material YBa$_2$Cu$_3$O$_7$ \cite{wu}.  Since then, several classes of these materials have been 
discovered (Fig.~\ref{cuprate-CS}), with one variant having a T$_c$ (under pressure) of 164 K \cite{gao}.

The discovery of the cuprates was a tremendous surprise.  It violated most of Matthias' rules, as it was a quasi-2D doped insulating oxide.
But the discoverers were not searching blindly.  Low temperature superconductivity had been seen many years before in the doped perovskite
SrTiO$_3$ at ridiculously low carrier concentrations \cite{schooley}.  Bednorz and Muller's guiding principle was to look at other oxides where Jahn-Teller
distortions played a crucial role \cite{rmp}.  The cuprates were a prime example, where such distortions lead to a half-filled d$_{x^2-y^2}$ level, 
copper being in a d$^9$ configuration.  This guidance was based on the idea that such strong lattice distortions could lead to strong coupling
electron-phonon pairing via bipolaron formation \cite{ranninger}.

But this bipolaron picture has turned out to be the minority view.  In fact, the community working on heavy fermions rapidly turned their attention
to the cuprates in 1987.  Simply reducing the dimensionality from three to two (a square lattice network) in theories based on antiferromagnetic spin
fluctuations led to the early prediction of $d_{x^2-y^2}$ pairing \cite{bickers} (Fig.~\ref{SF-pairing}).  Further support for this theory came out at about 
the same time when
neutron scattering revealed that the undoped parent insulating phase was a commensurate antiferromagnet with $Q=(\pi,\pi)$ \cite{vaknin}.  In the
beginning, there was a lot of resistance to such a non s-wave state, given the high T$_c$ and seeming insensitivity to disorder, but based on
the d-wave prediction, evidence begin to emerge that order parameter nodes were indeed present - penetration depth measurements indicated a linear in $T$
penetration depth at low temperatures \cite{bonn} and angle resolved photoemission was consistent with a node along the zone diagonal as 
expected for such a d-wave state \cite{shen}.  Definitive evidence came when phase sensitive Josephson tunneling was able to detect the
sign change in the order parameter upon ninety degree rotation \cite{wohlman}, at which point all but a few skeptics were convinced.

\begin{figure}
\includegraphics[width=3in]{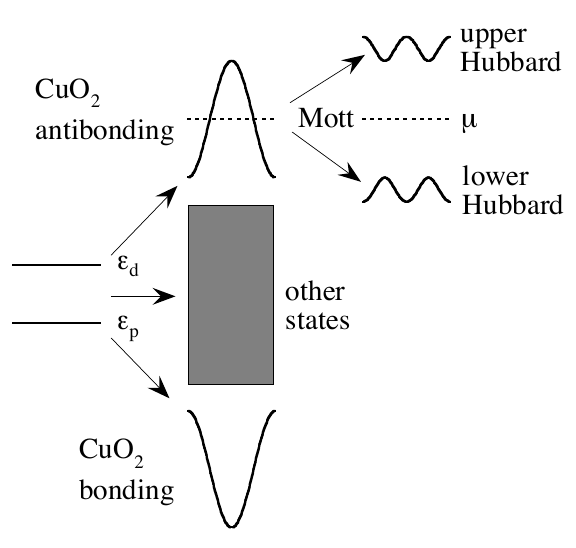}
\caption{Basic electronic structure of the cuprates.  Copper d$_{x^2-y^2}$ orbitals hybridize
with planar oxygen p$_x$ and p$_y$ orbitals, forming bonding and antibonding combinations.
Correlations cause the half filled antibonding band to split into lower and upper
Hubbard bands.}
\label{cuprate-es}
\end{figure}

One might have thought this would settle the debate, but such was far from the case.  In the same month that the discovery of YBCO was
announced, Phil Anderson proposed an alternative picture \cite{rvb}.  Anderson early on had realized certain crucial aspects of the cuprate problem - low
dimensionality, quantum limit of the spins (the single d hole has a spin of 1/2), and the nature of the insulating state (Fig.~\ref{cuprate-es}).  For a half filled band, band
theory predicts metallic behavior, but in the presence of a large Coulomb repulsion, $U$, the electrons would localize, forming a Mott insulating
state with an energy gap between a lower Hubbard band and an upper Hubbard band.  Band theory 
can simulate this by mapping one band to
an `up' spin state, and the other to a `down' spin state, but Anderson felt these arguments were fallacious, since they equated the exchange interaction
with the on-site $U$.  In fact, he felt that the Mott phenomenon was independent of whether the ground state was magnetic or not, and the theory
he proposed was based on an earlier paper seeking to understand the nature of frustrated magnetism \cite{phil73}.

\begin{figure}
\includegraphics[width=1.5in]{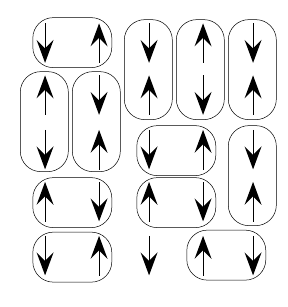}
\caption{An RVB state is a liquid of spin singlets, with unpaired spins denoted as spinons.}
\label{rvb}
\end{figure}

To understand this, note that in the Mott case, magnetism is induced by the superexchange interaction \cite{SX}.  In essence, if the spins are aligned between
near neighbors, there is no gain in the free energy by the Pauli exclusion principle, but if they are anti-aligned, one can gain energy by virtual hopping.
By second order perturbation theory, this energy is 4$t^2/U$ where $t$ is the hopping integral, defining the superexchange $J$.  Now consider
a N\'eel lattice.  The exchange energy for a given site is z$J$ where $z$ is the number of neighbors.  On the other hand, consider singlet formation
for $S=1/2$ spins.  The exchange energy per singlet is 3$J$.  For a square lattice, $z=4$, so the N\'eel state wins.  But allow the singlets to fluctuate
from bond to bond.  Anderson speculated that the resulting free energy gain might be sufficient to tip the balance in favor of a liquid of spin singlets
rather than a N\'eel lattice (Fig.~\ref{rvb}), hence the term `resonating valence bonds' (RVB).  Although Anderson was wrong in that the undoped material does form
a N\'eel lattice (but with a moment reduced to 2/3 its classical value), it was later found that for hole doped materials, only a few percent of holes is
sufficient to destroy magnetism, indicating that the basic idea might be right.

The RVB theory has been controversial to say the least.  One well known physicist quipped that the initials actually stood for `rather vague bullshit'.
Another wrote an extended poem (based on Hiawatha!) claiming Anderson was leading young physicists down the primrose path, supposedly to
their ultimate destruction \cite{hiawatha}.  Still, its profound influence in the field cannot be denied.

\begin{figure}
\includegraphics[width=3in]{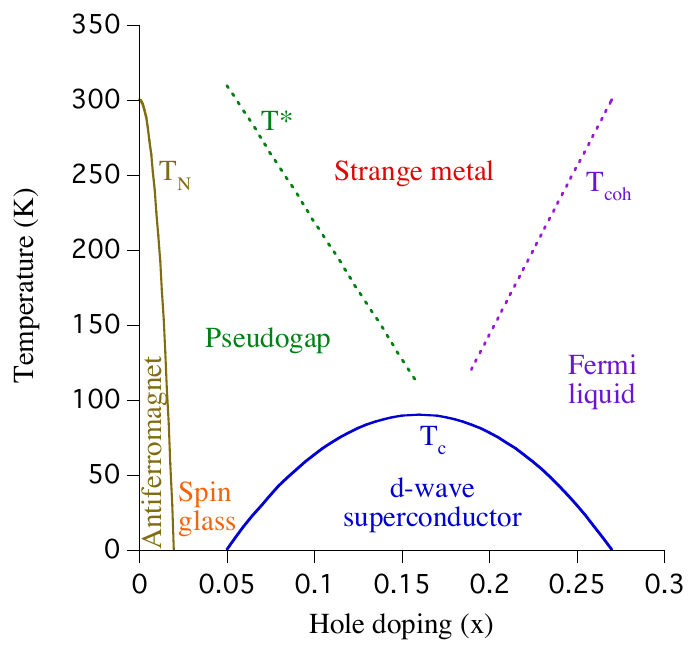}
\caption{Phase diagram of cuprates versus hole doping.  Three normal phases surround the
superconducting dome:  the pseudogap phase, and two gapless phases - a strange
metal exhibiting a linear $T$ resistance, and a more conventional Fermi liquid.}
\label{cuprate}
\end{figure}

\begin{figure}
\includegraphics[width=3.4in]{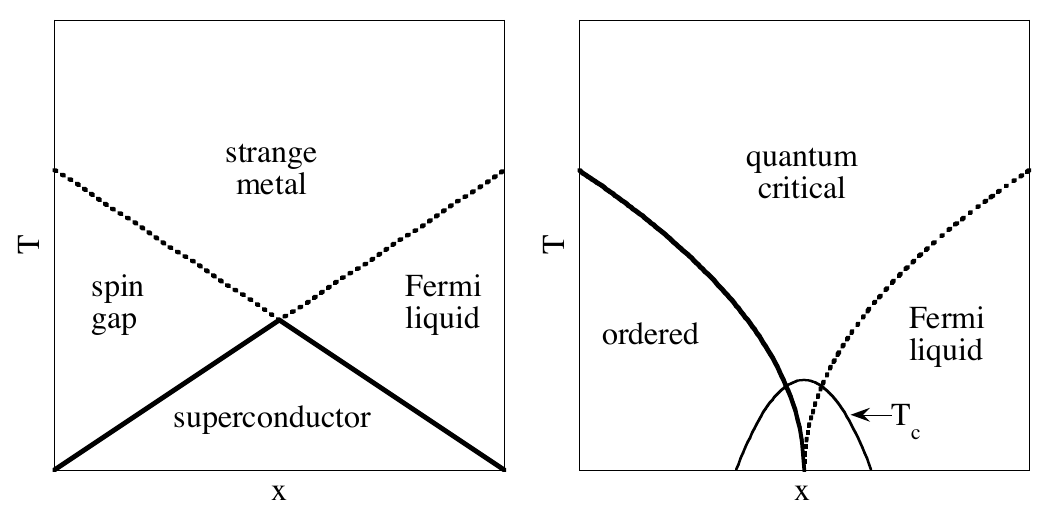}
\caption{Two proposed phase diagrams of the cuprates - RVB \cite{nag-lee} (left) and quantum critical \cite{varma00}
(right).}
\label{phase}
\end{figure}

Anderson's original theory was the so-called `uniform' RVB state.  In such a theory, free S=1/2 degrees of freedom (`spinons') form a Fermi surface.
But shortly afterwards, it was realized that upon doping with carriers (`holons'), the lowest energy ground state was equivalent to a d-wave liquid of
spin singlets \cite{kotliar}.  In fact, RVB theory gave one of the first predictions of the 
temperature-doping phase diagram of the cuprates (Fig.~\ref{cuprate}), with four regions identified (Fig.~\ref{phase}a).  Below a temperature 
T$^*$ that decreases
linearly with the doping, the d-wave spin liquid would form, leading to a d-wave energy gap in the spin excitation spectrum.  Below a temperature
T$_{coh}$ that increases linearly with the doping, the charge degrees of freedom would become phase coherent, leading to Fermi liquid behavior.
Below both temperatures, the combination of a d-wave spin singlet with charge coherence would give rise to a d-wave superconductor, which thus
forms a `dome' in the temperature-doping phase plane.  Above both temperatures, one would have instead a `strange metal' phase, exhibiting
gapless non Fermi liquid behavior.  There are some photoemission data which are in support of this picture for the phase diagram \cite{utpal11}.

\begin{figure}
\includegraphics[width=1.5in]{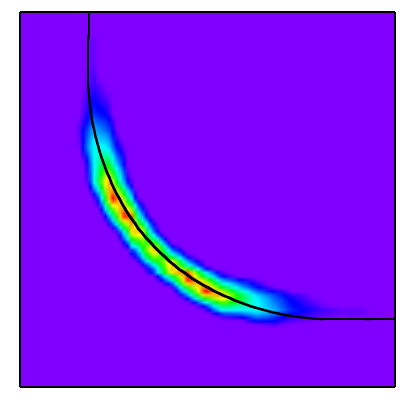}
\caption{Zero energy intensity from photoemission for the cuprate Bi2212 in the pseudogap phase,
exhibiting an arc of gapless excitations \cite{norm07}.  The large Fermi surface in the gapless normal phase is shown as the black curve.}
\label{arcs}
\end{figure}

These ideas were emerging at about the same time as NMR experiments were revealing the presence of a `spin gap' that roughly had the doping
dependence indicated by the RVB theory \cite{alloul}.  Subsequently, this `pseudogap' was revealed by a number of other probes, including
c-axis infrared conductivity \cite{homes}, photoemission \cite{marshall,ding,loeser} and tunneling \cite{renner}.  Its observation by angle
resolved photoemission (ARPES)
was particularly illuminating, in that the inferred gap appeared to be d-wave like in nature.  How d-wave like is a matter of continuing debate.
What is clear is that the Fermi surface is truncated in the pseudogap phase into `arcs' centered at the nodes of the d-wave superconducting
state \cite{arcs,kanigel} (Fig.~\ref{arcs}).  What is not clear yet is whether these arcs represent one side of a closed pocket in momentum 
space \cite{marshall,yang1,yang2} or
a thermally broadened d-wave node \cite{norm07,reber}.  The latter is consistent with RVB theory, and further evidence has been given
by its consistency with some low temperature photoemission data for non superconducting samples \cite{utpal10} which continue to exhibit
a d-wave like gap.  But increasing attention has been given to the former possibility.

If some kind of order were present in the pseudogap phase, a reconstruction of the Fermi surface into smaller pockets would be expected.
For instance, simple N\'eel antiferromagnetism in the doped case would initially give rise to a small hole pocket centered around
the $(\pi/2,\pi/2)$ points \cite{chub-morr}.  In the early days of cuprates, such a possibility was actively discussed, and was implied as well in
the initial ARPES study of Marshall \etal~\cite{marshall}.  The idea here is that the transition to long range magnetic order is determined by coupling
between the CuO$_2$ planes, since Heisenberg spins in two dimensions do not order.  As mentioned above, a few percent of doped holes is sufficient
to disrupt this order.  Still, fluctuating two dimensional order is likely still present, and if the fluctuations are slow enough, an apparent pocket might
be formed \cite{kampf}.  The resulting `shadow' bands were subsequently seen by several ARPES studies \cite{aebi}, but in all cases we know, they appear
to actually be due to the crystal structure - Bi2201, Bi2212, and LSCO have $(\pi,\pi)$ as a reciprocal lattice vector due to orthorhombic distortion of the
crystal lattice.

This picture, though, got further support when quantum oscillation data finally emerged.  In the early days of cuprates, such studies were done, but
led to inconclusive results.  But with the advent of high quality crystals, the first definitive data appeared in 2007 \cite{louis1}.  These initial experiments
were done on underdoped YBCO (the so-called ortho-II phase with a well ordered crystal structure).  What they revealed was a small pocket,
first seen by quantum oscillations of the Hall resistance.  But interestingly, the Hall resistance was negative, indicating that the pocket
was an electron pocket, despite the fact that one is hole doping \cite{louis2}.  This led to the speculation that such pockets could arise from
incommensurate order due to the formation of magnetic stripes \cite{millis}.

\begin{figure}
\includegraphics[width=3in]{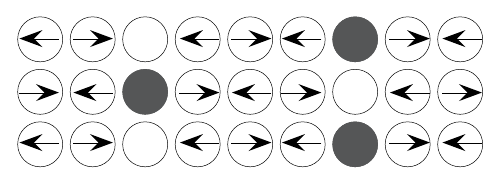}
\caption{A schematic for stripes, where doped holes (dark circles) form ribbons of charge
separated by undoped antiferromagnetic regions \cite{tranquada}.}
\label{stripes}
\end{figure}

Such magnetic stripes were first identified by neutron scattering \cite{tranquada} (Fig.~\ref{stripes}).  They are particularly pronounced near 1/8 doping.
There are two
ways one might think of such stripes.  First, as an incommensurate spin density wave state, similar to chromium.  Here, the incommensurability is
due to doping, which moves the chemical potential away from half filling for the hybridized copper-oxide band.  The other picture is a real space one - doped
holes do not go in homogeneously, but in order to minimize their Coulomb repulsion, form rivers of charge \cite{zaanen,emery-K}.  In between these
rivers of charge are undoped antiferromagnetic regions.  Therefore, the `incommensurability' in this case is due to a phase slip of the simple N\'eel
lattice when moving across the stripes.  The lack of observation of higher harmonics in the neutron data seemed to suggest the former, but spectacular scanning
tunneling data seem most consistent with the real space picture \cite{steve-rmp}.  The fact that quantum oscillation data and the region of negative
Hall effect seem to form a dome around 1/8 doping definitely point to stripes as the origin of the pockets \cite{hall}.

The remaining question has concerned charge versus magnetic stripes.  In 1/8 doped LBCO (one of the few materials where static stripe order
is observed), charge ordering occurs before spin ordering \cite{john2}.  Charge ordering as an explanation of the quantum oscillation results
had been discounted because of difficulties in getting an electron pocket in that case \cite{millis}, but it was subsequently shown that a nematic
distortion (where x-y degeneracy is broken) was sufficient to stabilize them \cite{yao}.  In fact, one generally expects that as one reduces the 
temperature, nematicity appears first, followed by charge order and then eventually by spin order \cite{steve-rmp,vojta}.  Interestingly, data on the Nernst
effect in YBCO are consistent with nematicity setting in at the pseudogap temperature, T$^*$ \cite{nernst}.  But the problem with these scenarios
is that the electron pocket is in the $(\pi,0)$ region of the Brillouin zone, exactly where ARPES sees a large pseudogap.

Because of this, an alternate picture has emerged \cite{harrison}.  Here, the Fermi arcs instead of closing towards the $(\pi/2,\pi/2)$ points (which would
form hole pockets) instead close towards the $(0,0)$ point (to form electron pockets).  The translation of the arcs to form such a pocket
 is achieved by having biaxial charge order.  Recently, such order has indeed been seen by x-ray studies \cite{keimer,hayden2}.  
 So, this would seem to settle matters, except for the fact that
no evidence for a closed pocket near the $(0,0)$ point of the Brillouin zone has ever been inferred from photoemission data.

To complicate matters, another type of novel magnetic order has been seen to set in at T$^*$ \cite{bourges}.  The origin of this finding goes back to the
early days of cuprates when it was realized that in the `strange metal' phase, the resistivity was linear in temperature \cite{martin}.
Although at high temperatures
this is not a surprise (the electron-phonon interaction can cause this), at lower temperatures this was a puzzle, particularly since it was observed
in samples of Bi2201 where T$_c$ was very low.  Although various models have been suggested to account for this linearity, the most
straightforward one was proposed by Varma and collaborators in 1989 \cite{varma89}.  If one has a bosonic spectrum that is flat in energy ($\omega$),
then the imaginary part of the fermion self-energy due to interaction with those bosons
will be linear in $\omega$.  If one assumes a momentum independent interaction, then this translates
to a linear T resistivity.  This has been denoted as marginal Fermi liquid theory.  The experimental motivation for this conjecture was the roughly
frequency independent background observed in Raman scattering.  Further support for this conjecture was found when a linear $\omega$
behavior of the imaginary part of the self-energy was identified by ARPES \cite{valla}.  A subsequent ARPES study was consistent with this
linear $\omega$ term being roughly momentum independent \cite{adam04}.

\begin{figure}
\includegraphics[width=3.4in]{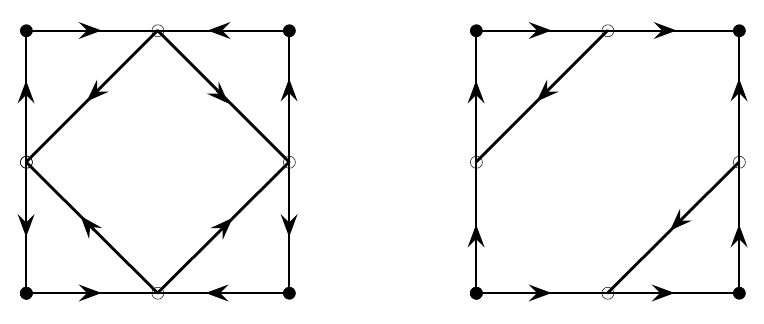}
\caption{Two orbital current patterns proposed by Varma \cite{simon}.  Filled circles are copper, empty
circles oxygen.  The right pattern is consistent with photoemission \cite{adam02} and neutron scattering \cite{bourges} data.}
\label{OC}
\end{figure}

Later, Varma proposed a microscopic theory along these lines \cite{varmaOC}.  His conjecture was that the single band Hubbard
model, which was the theoretical underpinning for most theories, was an inadequate model for the cuprates.  In particular, because of the
hybridization between the copper $d_{x^2-y^2}$ orbital and the oxygen $p_x$ and $p_y$ orbitals, he felt that a three band model was a minimal
description.  In the process of studying such a model, he found a new ground state where currents flowed inside the CuO$_2$ network of
ions.  Although flux states had been proposed before (they occur in RVB models), this flux state was unique in that it did not break translational
symmetry (Fig.~\ref{OC}).  In essence, it is an orbital antiferromagnet with $Q$ = 0 (allowed since there are two oxygens in the square lattice unit cell).
An initial neutron scattering study did not find this state \cite{lee}, but a subsequent ARPES study using circularly
polarized light potentially identified it via dichroism \cite{adam02}. This identification, though, required a rotation of the originally predicted
current pattern by 45$^{\circ}$
(an alternate ground state that Varma had not initially considered).  Once this was realized, neutron scattering indeed identified the state in
underdoped YBCO \cite{bourges}.  Subsequent studies have found this state in Hg2201, Bi2212, and a short range ordered version in LSCO \cite{bourges-rev}.

The observed moment is substantial, up to a few tenths of a $\mu_B$ per CuO$_2$ unit.  But it has not been observed by either NMR or $\mu$SR
measurements \cite{NMR-OC}.
This has led to some skepticism that the effect could be an artifact - structural transitions can lead to a change in the spin flip
ratio in neutron scattering, and such a structural effect could explain the ARPES dichroism results as well.  If it were some novel structural
transition, though, it has yet to be identified, though there is a claim of seeing inversion breaking from x-ray natural 
dichroism \cite{kubota}, which would also be consistent with recent STM results where a difference is seen between the two oxygen 
sites \cite{STM-nematic}.  Whatever it is, it does have an order parameter
like evolution that sets in at T$^*$, which confirms Varma's original conjecture that the pseudogap phase represents some sort of symmetry breaking.

Subsequent work by Kapitulnik's group identified an optical Kerr rotation that sets in below the T$^*$ line identified by neutron scattering in YBCO \cite{kerr},
but appears to be coincident with it in Bi2201 \cite{he}.  It now appears that the Kerr signal is coincident with the biaxial charge order recently
identified in underdoped YBCO \cite{keimer,hayden2}.  This has led to speculation that the Kerr signal might be due to some kind of `chiral' charge 
density wave that breaks inversion symmetry \cite{hosur} (in Bi2201, the crystal space group already breaks inversion symmetry).  Alternately,
the Kerr signal could simply be a signature of a magneto-electric phase, as occurs in antiferromagnetic Cr$_2$O$_3$ \cite{orenstein} or in an orbital
current phase with a structural distortion as Varma suggests \cite{varma-new}.  Whether these various symmetry breakings can explain a large pseudogap 
remains to be 
seen.  Certainly, the T dependence of the pseudogap identified by ARPES follows that of the Kerr signal in Bi2201 \cite{he}.
But nematics, which orbital currents are related to, do
not necessarily generate an energy gap, and it is also doubtful whether the weak charge order identified by x-ray scattering could cause a large energy
gap.  Stripe models, on the other hand, do generate a gap, with the spin gap in the undoped regions between the stripes inducing a gap in
the mobile holes from virtual hopping of the holes into these regions \cite{carlson}.  Virtual hopping of pairs of mobile holes into these undoped regions is also
a potential source for the superconductivity \cite{carlson}.

One reason for highlighting all of these results (nematics, stripes, orbital currents, etc.) is not only to emphasize the complexity of the pseudogap phase,
but that such results highlight the strong possibility of an alternative phase diagram to the RVB one, where an ordered phase is suppressed to zero
by doping, ending at a quantum critical point (Fig.~\ref{phase}b).  The T$_{coh}$ phase line would then be the `quantum disordered' mirror of the T$^*$ line.
Above these two lines, quantum criticality would occur, which would then explain the non-Fermi liquid behavior of the `strange metal' phase.
More importantly, if we make an analogy to the previous section on heavy fermions, one might suspect that the fluctuations in the quantum critical
regime associated with the pseudogap phase would be the origin of the pairing in the superconducting phase.  Regardless, since superconductivity
is an instability of the normal phase, and the fact that over much of the phase diagram, superconductivity occurs below the $T^*$ line, a proper identification 
of the nature of the pseudogap phase will be critical for the ultimate theory of cuprates \cite{AP}.

\begin{figure}
\includegraphics[width=3.4in]{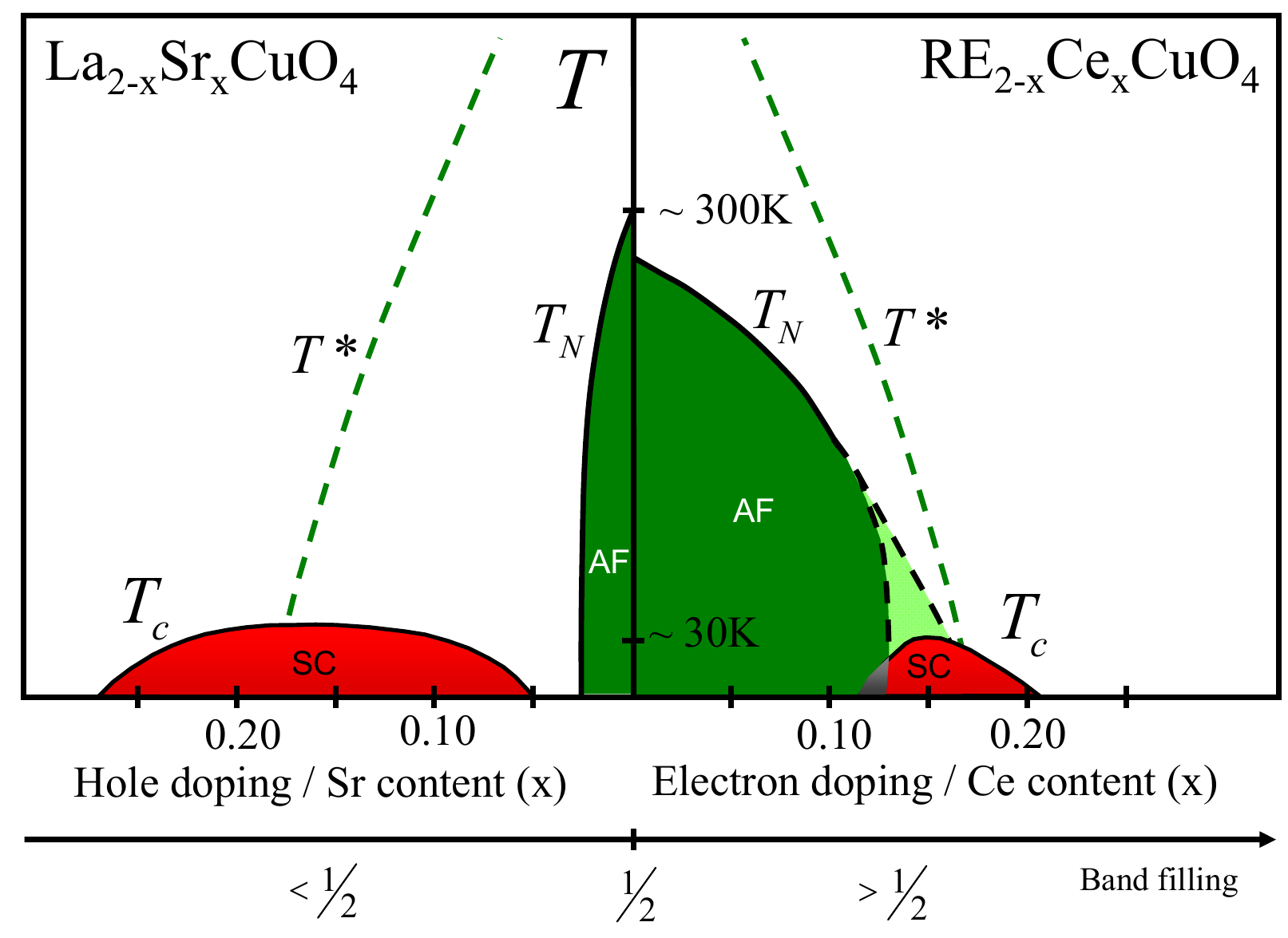}
\caption{Combined electron and hole doped phase diagram of the cuprates \cite{greene-RMP}.  Note that
the antiferromagnetic phase extends over a larger doping range in the electron doped
case.}
\label{ehphase}
\end{figure}

Superconductivity also occurs in electron doped cuprates \cite{greene-RMP} (Fig.~\ref{ehphase}).  Here, commensurate antiferromagnetism occurs
over a much larger range of doping than in the hole-doped case, with the pseudogap phase associated with this magnetism (in the 2D limit, one 
expects a pseudogap phase in the renormalized classical regime above the magnetic ordering temperature \cite{vilk}).  Many of the properties of the electron
doped side are similar to the hole doped one (d-wave superconductivity, pseudogap, non Fermi liquid behavior), suggesting that the origin of
superconductivity is the same.  If so, this is definite support for those theories which suggest that magnetic correlations are responsible for the
pairing, as thought to be the case in heavy fermions.  Although it was originally felt that magnetic correlations weaken significantly with doping, recent resonant
inelastic x-ray (RIXS) studies on YBCO \cite{letacon} and Bi2212 \cite{dean} indicate that strong spin fluctuations are still present at optimal doping.

The various theories discussed above have led to a passionate debate on the nature of the pairing in cuprates.  In antiferromagnetic spin fluctuation
theories, pairing is treated in an approximation similar to the electron-phonon case - that is, by virtual exchange of spin 
fluctuations (Fig.~3) \cite{SLH,MSV,BBE,SLH2}.  The pairing interaction is proportional to the dynamic spin susceptibility, and thus the source of pairing
is an induced interaction that is confined to energies of order 0.4 eV or less.  This is in contrast to RVB theories, where the pairing is encoded in
the `normal state' wavefunction, and the interaction is associated with the superexchange $J$ which should only develop dynamics on an energy
scale of order $U$ \cite{glue}.  Dynamical mean field theory calculations in the cluster approximation are in support of the former picture \cite{maier}
even though such calculations do exhibit RVB like behavior, with singlet formation particularly pronounced for four site copper 
plaquettes \cite{haule-P}.  Certainly, changes
in the optical response of cuprates below T$_c$ have been observed up to 5 eV \cite{rubhausen} indicating that the effects of pairing extend over
a large energy range.  This may be related to other optics experiments \cite{vdM} that indicate a lowering of the kinetic energy below T$_c$
in underdoped materials, where the resulting increase in low energy spectral weight would come at the expense of high energy spectral weight
(coming from an energy scale of $U$).  This is very different from the increase of the kinetic energy that occurs in BCS theory due to particle-hole mixing.
In essence, the potential energy decreases when the energy gap is formed in the pseudogap phase, but the electrons remain incoherent.
Only below T$_c$ does coherence occur, leading to a decrease in the kinetic energy.  This has been suggested to be in support of pre-formed
pairs in the pseudogap phase (as also implied by the large Nernst signal in the pseudogap phase \cite{ong}), but kinetic energy lowering has been seen
as well in dynamical mean field calculations where the existence of pairing above T$_c$ has not been identified \cite{DMFT-kinetic}.

Given the diverse nature of the phenomena in cuprates, it has been difficult to come up with a `smoking gun' for pairing.  Attempts to extract the 
anomalous
self-energy from planar tunneling, ARPES, and scanning tunneling probes have been inconclusive up to now, mainly because of the strong
momentum dependence associated with d-wave pairing, along with the complications of a normal state pseudogap, though looking at the
angle resolved density of states instead can help \cite{vekther}.
Attempts to analyze the `normal state' self-energy indicate the presence of spin fluctuations \cite{shen97,norm97}, phonons \cite{lanzara},
and a frequency independent bosonic background similar to what is seen in Raman scattering \cite{eschrig,zhou-V}.  Much focus has been
put on the `spin' resonance below T$_c$, which was first identified in cuprates \cite{rossat} before being seen in several heavy fermion superconductors
(and later in pnictides).  Although this could simply be consistent with having d-wave pairing (the d-wave state reverses sign under translation
by $Q=(\pi,\pi)$), neutron scattering does indicate that the formation of the resonance is associated with a lowering of the overall exchange
energy below T$_c$ \cite{dai}, though it should be remarked that because of phonon contamination in the data, uniquely extracting the spin fluctuations
over a large range of energy and momentum is difficult.  Certainly, phonons have been argued to play a large role in the normal state self-energy \cite{lanzara},
particularly at low dopings where polaronic effects are evident \cite{eugene}, but it is a stretch to believe that phonons are responsible for d-wave
pairing at the high temperatures observed in the cuprates, though some have advocated this \cite{nagaosa}.

What should be remembered is that
ARPES for overdoped materials (where the complications of a pseudogap are not present) is consistent with an energy gap of the functional
form $\cos(k_xa)-\cos(k_ya)$ \cite{shen,ding-gap}.  This implies pairing originating from near neighbor copper interactions (since this function is the Fourier 
transform of such).  It is doubtful whether phonons would give rise to this particular functional form - or intra-unit cell orbital currents for that matter,
where the pairing vertex is of the form $({\bf k} \times {\bf k'})^2$ \cite{aji}.  Spin
fluctuations can, as well as RVB theories.  It has been argued that these latter two approaches represent opposite limits of a more general theory \cite{bob},
but Anderson has argued against this \cite{phil-bob}.  Certainly, there is a difference between local singlets (RVB) as opposed to longer range
antiferromagnetic spin fluctuations.  Regardless, the real worry is that as in $^3$He, everything and the kitchen sink might be
contributing to the pairing.

Ultimately, it may take unbiased numerical approaches to settle these matters.  Quantum Monte Carlo (QMC) simulations of fermionic systems suffer from 
the sign problem where negative probabilities occur, meaning that one is limited in how low in temperature one can do reliable calculations.
Most QMC simulations of the single band Hubbard and $t-J$ models do indicate d-wave superconductivity \cite{sorella}.  Another essentially
exact approach is the density matrix renormalization group (DMRG) approach \cite{white}, but extending this to two dimensions requires simulating finite width
strips \cite{white2}.  Such simulations show stripe formation as well as d-wave pairing \cite{white3,PEPS}
Approximations to DMRG have been developed for 2D, including PEPS (projected entangled pair states) \cite{PEPS} and MERA (multiscale
entanglement renormalization ansatz) \cite{MERA}, which attempt to preserve certain correlations during coarse graining.  PEPS simulations have been
particularly illuminating, finding near degeneracy of striped and paired states \cite{PEPS}.

Perhaps the most popular approach has been dynamical mean field theory (DMFT) \cite{DMFT} and its various cluster extensions
(either in momentum space or real space).  This involves attempting to do a solution of the `exact' problem for a cluster, and then embedding
this cluster in a bath, with the bath-cluster interaction treated by hybridization as in an Anderson impurity model, in order to represent the full 
periodic system.  Typically,
a quantum Monte Carlo solver is used, again limiting one in the temperature range that can be accessed, though ironically, this is less a restriction
for pairing since the bosonic nature of the pair state somewhat ameliorates the sign problem \cite{gull1}.  Clusters up to 16 sites have been treated, though
it will take larger clusters to verify convergence in regards to symmetry breaking ground states such as magnetism or superconductivity.
As mentioned above, four site clusters are consistent with singlet formation \'a la RVB \cite{haule-P},
though it is now recognized that such clusters overemphasize singlet formation.
Still, DMFT methods have evolved to the point where they can now explain quantitative trends in the cuprates, such as the variation
of T$_c$ with various on-site and hopping energies, including the important role of the apical oxygens \cite{gabe-EPL}.

The most recent DMFT results indicate that the pseudogap is a precursor of the Mott insulating gap at zero doping, and as it is suppressed,
superconductivity appears.  Since this gap primarily affects states near $(\pi,0)$ (antinodal states), the unusual nodal-antinodal dichotomy
revealed by photoemisson, where nodal states are gapless and coherent, and antinodal states gapped and incoherent, is naturally
explained \cite{N-AN}.  This gives new insight into the nature of the Fermi `arcs', and follows earlier speculations by Bob Laughlin that the pseudogap seen
in ARPES extrapolates to the Mott gap as the doping is reduced \cite{bob-ARPES}.  This approach is also in line with the basic RVB idea
that the large superexchange $J$ that is a unique signature of cuprates is the source of pairing, though again, detailed calculations of
the anomalous self-energy give results more reminiscent of spin fluctuation theory \cite{maier}.  In that context, it should be remarked that RVB is usually
presented in a `mean field' approximation.  One approach to go beyond this is by including gauge fluctuations (to capture the constraint of no double
occupation) which does introduce significant low energy dynamics \cite{lee-RMP}, but whether this is a controlled approximation is unclear.

Regardless, it appears that magnetic correlations of some sort are responsible for d-wave superconductivity in the cuprates.  Whether this should best
be thought of as singlets, paramagnons, orbital currents, or a combination thereof remains to be seen.  Mott physics certainly plays a role, though
it should be remarked that overdoped cuprates emerge from a more or less normal Fermi liquid phase.  But even if this is so, describing the wealth
of phenomena that has been revealed by such techniques as angle resolved photoemission, neutron and x-ray scattering, and scanning tunneling 
microscopy will keep researchers busy for many years to come.

\section{Organic Superconductors}

Although most organics are insulators, some can be metallic.  Interest in the possibility of superconductivity goes back to 1964 when Bill Little
proposed that such materials could be high temperature superconductors \cite{little}.  This promoted a flurry of activity, including even a conference,
leading Bernd Matthias to once quip that this was the first one he knew that was devoted to non-existent materials \cite{bernd}.  But in 1980, the real
deal was reported by Denis Jerome's group in a quasi-1D Bechgaard salt \cite{jerome}, followed up by its discovery in quasi-2D
variants \cite{greene,williams,jerome2}.  A nice review of this field recently appeared \cite{ardavan}.

A typical example of the quasi-1D variant is (TMTSF)$_2$PF$_6$.  At ambient pressure it exhibits a spin density wave, probably due to Fermi
surface nesting, that onsets at about 12K.
Under pressure, the SDW is suppressed, after which superconductivity appears at about 1K.  These materials exhibit upper critical fields far in
excess of the Pauli limiting field, indicating (at least at high fields) that the pairing is triplet in nature.
Similarities to cuprates has been emphasized in recent work \cite{taillefer}.

\begin{figure}
\includegraphics[width=3in]{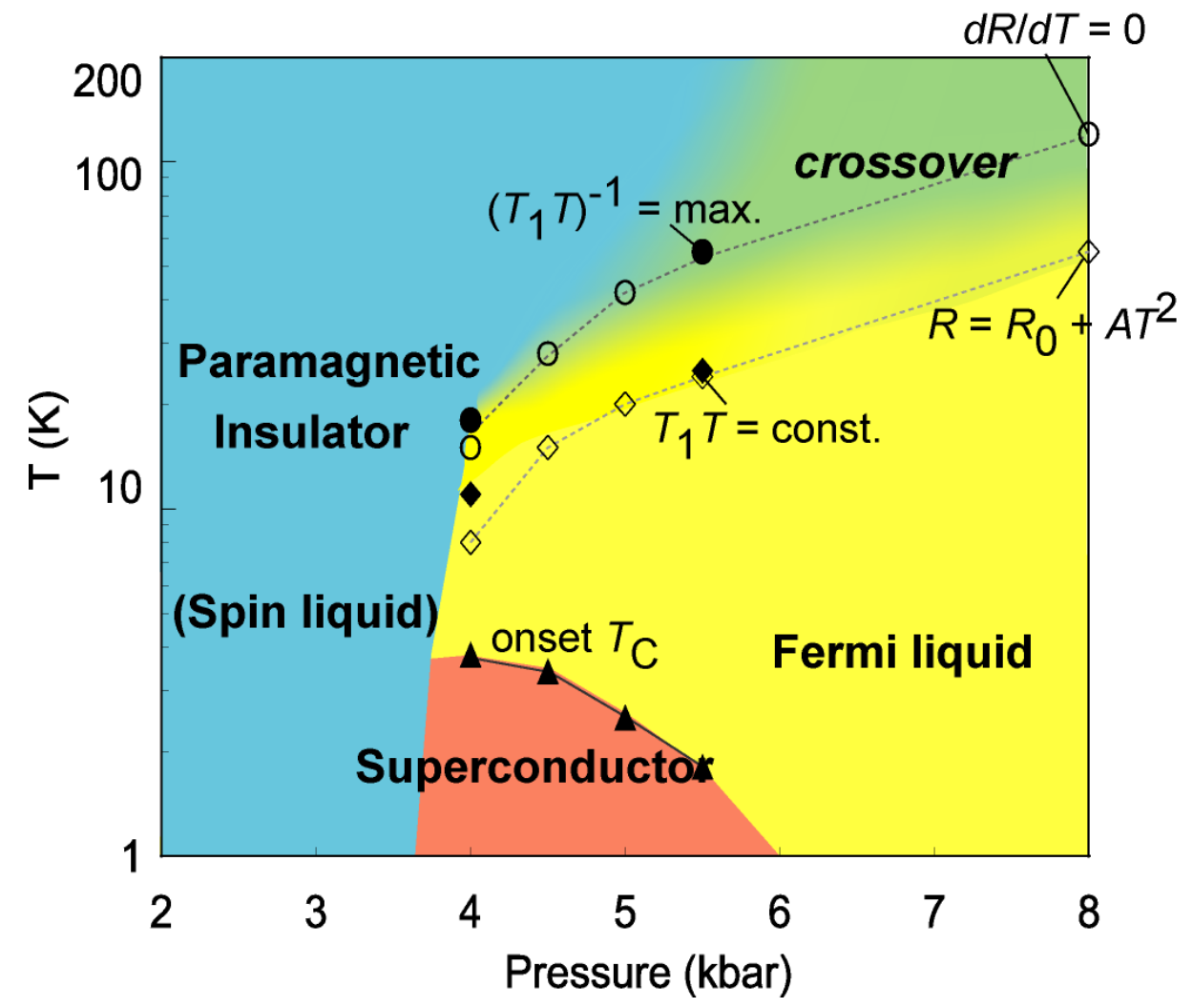}
\caption{Phase diagram of $\kappa$-(ET)$_2$Cu$_2$(CN)$_3$ versus pressure \cite{kurosaki}.
A superconducting phase abuts a Mott insulating phase with no long range magnetic
order.}
\label{organic}
\end{figure}

Perhaps of more interest are the quasi-2D variants, which exhibit superconductivity up to 13K.  Typically, these materials are composed of molecular
dimers which form a triangular lattice, with one spin 1/2 degree of freedom per dimer.  Such frustrated lattices were the original source of inspiration
for Anderson's RVB theory \cite{phil73}.  Based on this, there has been a lot of interest in the phase diagram of these materials.  Of recent
relevance is $\kappa$-(ET)$_2$Cu$_2$(CN)$_3$ \cite{kurosaki}.  At ambient pressure, the material appears to be a Mott
insulator but with no evidence for long range magnetic order, implying the ground state is a spin liquid.  Low temperature
specific heat measurements
are consistent with the presence of a Fermi surface \cite{thermalC}, perhaps the long sought after `spinon' Fermi surface of RVB lore \cite{leeNV},
though it should be remarked that a transition of unknown origin has been detected at 6K by thermal expansion \cite{manna}.
Under pressure, a superconducting phase appears \cite{geiser} whose maximum T$_c$ abuts the spin liquid phase (Fig.~\ref{organic}).
This phase is reminiscent of that seen in underdoped cuprates, with a pseudogap effect apparent in NMR data \cite{ardavan} along with
an enhanced Nernst signal above T$_c$ \cite{nam}.

Little is known about the gap structure of the organics.  The NMR relaxation rate  varies as T$^3$ suggestive of nodes (as in the cuprates),
though it should be remarked that this T$^3$ behavior seems ubiquitous in many materials regardless of their nodal structure.   Recently,
there has been some success with photoemission in this class of compounds \cite{kiss}, so it is hoped in the near future that more definitive 
evidence of the nature
of the superconducting state will be forthcoming.  Certainly, the available evidence points to a strongly correlated state, where Mott physics \cite{ross}
and magnetic correlations play a fundamental role, implying these materials are close cousins of the cuprates.

\begin{figure}
\includegraphics[width=3in]{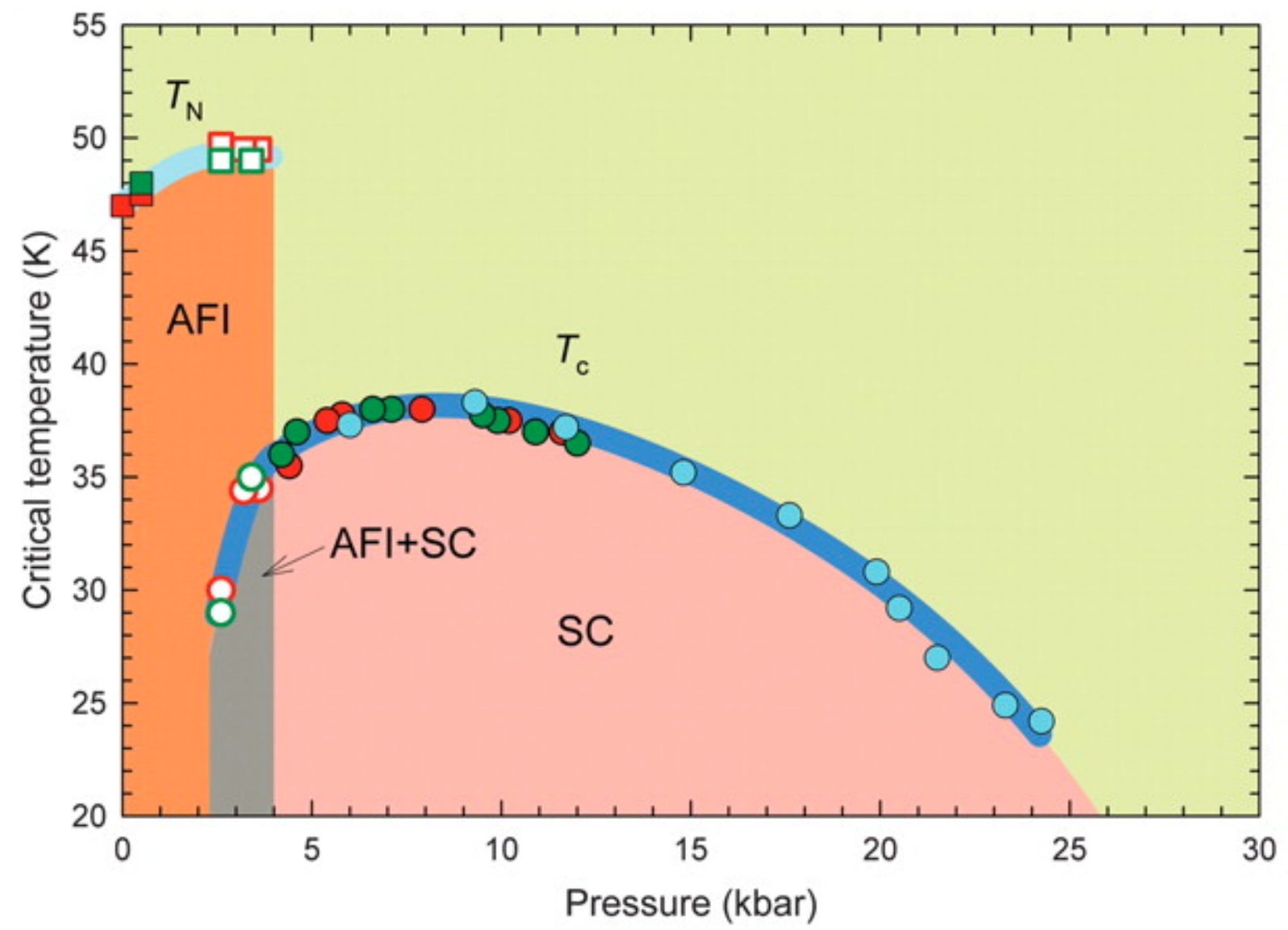}
\caption{Phase diagram of Cs$_3$C$_{60}$ versus pressure \cite{taka}.  Note the presence of
an antiferromagnetic insulating phase as in the cuprates.}
\label{bucky-PD}
\end{figure}

Besides these materials, a variety of other organic compounds have been discovered which are superconducting.  Of particular interest are
buckeyballs (C$_{60}$), which when doped with alkali atoms exhibit superconductivity up to 40 K \cite{bucky}.  For a long time, these were
regarded as strong-coupling conventional superconductors, but recent work on the 40 K cesium variety \cite{taka}
indicates a phase diagram again reminiscent of the cuprates and ET salts, where superconductivity emerges under pressure from
an antiferromagnetic insulating phase (Fig.~\ref{bucky-PD}).  Even more recently, high temperature superconductivity has been reported
in materials based on chains of benzene rings with superconductivity up to 33 K \cite{kobo,xue}.  More work will be necessary in order
to understand the relation of these materials to the organic salts described above.

\section{Pnictides}

\begin{figure}
\includegraphics[width=3.4in]{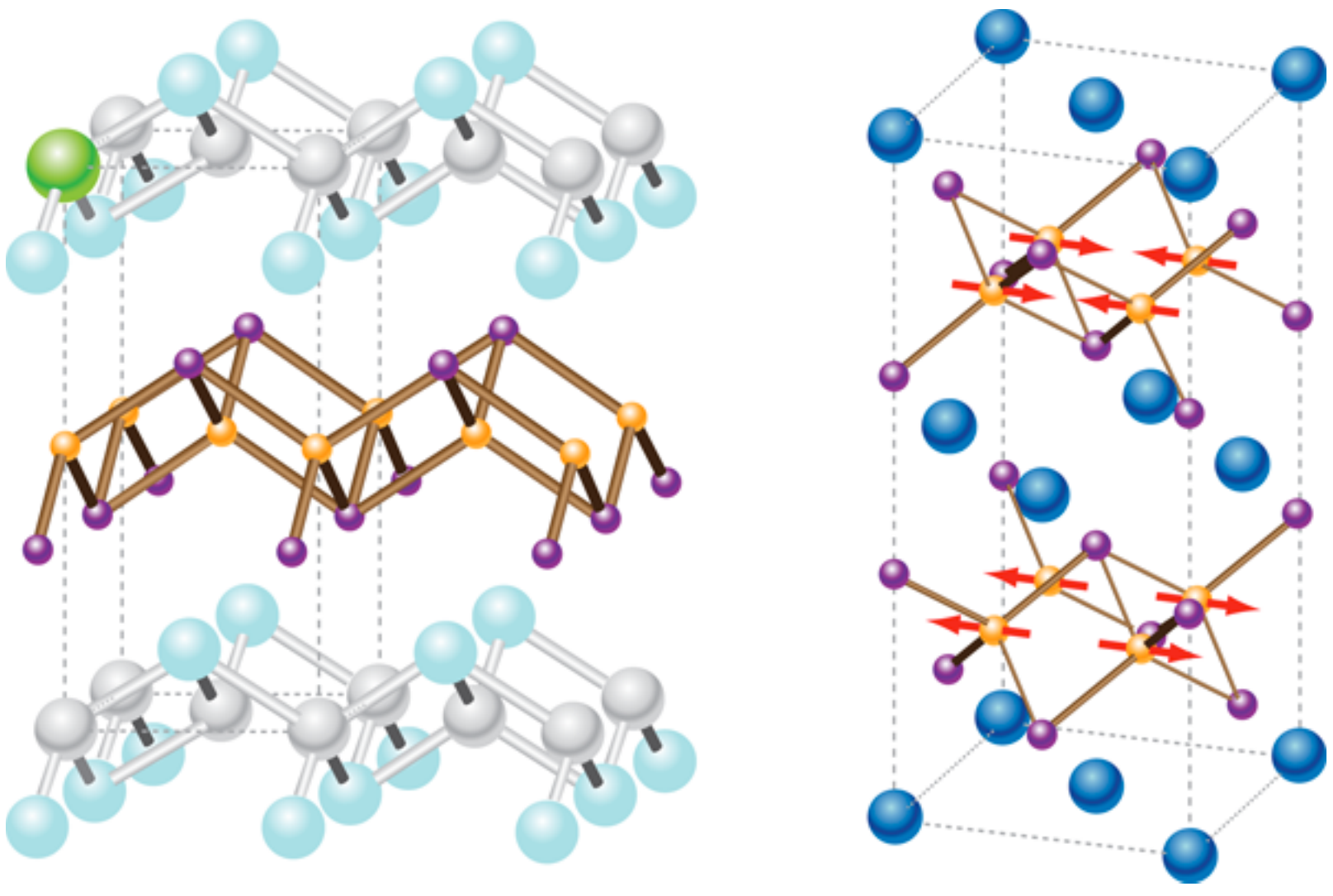}
\caption{LaOFeAs (left \cite{takahashi}) and CaFe$_2$As$_2$ (right \cite{goldman}) crystal structures, denoted as 1111 and 122 respectively.
Yellow are iron atoms, purple are arsenic ones.  On the left, a flourine dopant is shown in green.  On the right,
the spin directions (red arrows) are shown for the magnetic phase.}
\label{FeAs}
\end{figure}

In early 2008, Hosono's group announced the discovery of high temperature superconductivity in an iron arsenide compound \cite{hosono},
following earlier work by this group that had found lower temperature superconductivity in the phosphide variant.  Superconductivity was soon
seen up to 56 K \cite{56K}.  Several known crystal structure classes have now been identified (Fig.~\ref{FeAs}), the most studied being the
so-called 122 structure \cite{122} which has the same ThCr$_2$Si$_2$ structure as several heavy fermion superconductors.  The materials are 
composed of square lattices of iron atoms each tetrahedrally coordinated to arsenic ones, though the simpler `11' class of materials are 
actually iron chalcogenides.
FeSe has a relatively lower T$_c$ of 10K \cite{fese}, though an intercalated version has a T$_c$ above 40K \cite{fese-40K}.

\begin{figure}
\includegraphics[width=3in]{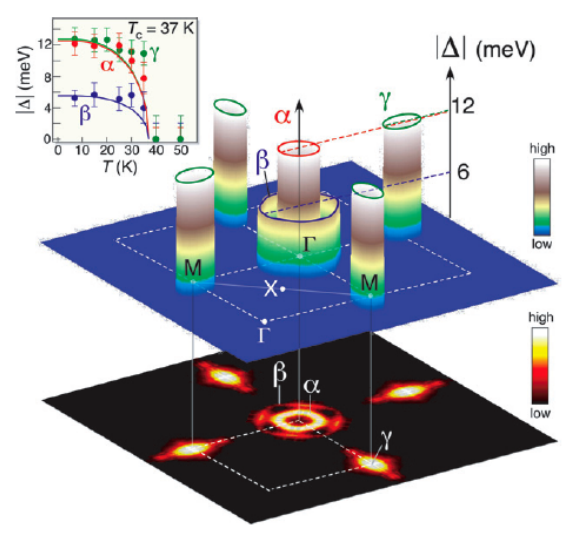}
\caption{Photoemission results for potassium doped BaFe$_2$As$_2$, with the superconducting
energy gap denoted as $|\Delta|$.  Two hole surfaces occur around $\Gamma$ and
an electron surface around $M$.}
\label{FeAs-PES}
\end{figure}

Like the cuprates, the undoped variant of the arsenides is a commensurate antiferromagnet \cite{cruz},
but in the pnictides it is metallic, though ARPES data reveal a Dirac-like dispersion
of the electronic states \cite{dirac} which is consistent with quantum oscillation studies \cite{QO-FeAs}.
This has led to the general feeling that the ground state is a spin density wave metal driven by Fermi surface
nesting as in chromium, though a more localized magnetic picture has been advocated by some \cite{si}.
Unlike the cuprates, where the spins form a checkerboard pattern, the magnetic order in pnictides is stripe-like instead \cite{cruz}.
This is consistent with `nesting' of the Fermi surface, which is composed of hole surfaces centered at the $\Gamma$ point and electron
surfaces centered at the zone edge ($M$ point), with the separation of these two centers equal to the magnetic ordering vector $Q$ \cite{FeAs-LDA}
(Fig.~\ref{FeAs-PES}).
The magnetic transition is associated with an orthorhombic distortion of the lattice which usually appears at a slightly higher temperature,
though in some materials they are coincident.  Sometimes the structural transition appears to be second order, other times first, but intriguingly,
a large nematic effect has been identified well above this transition \cite{nematic}.

\begin{figure}
\includegraphics[width=2.75in]{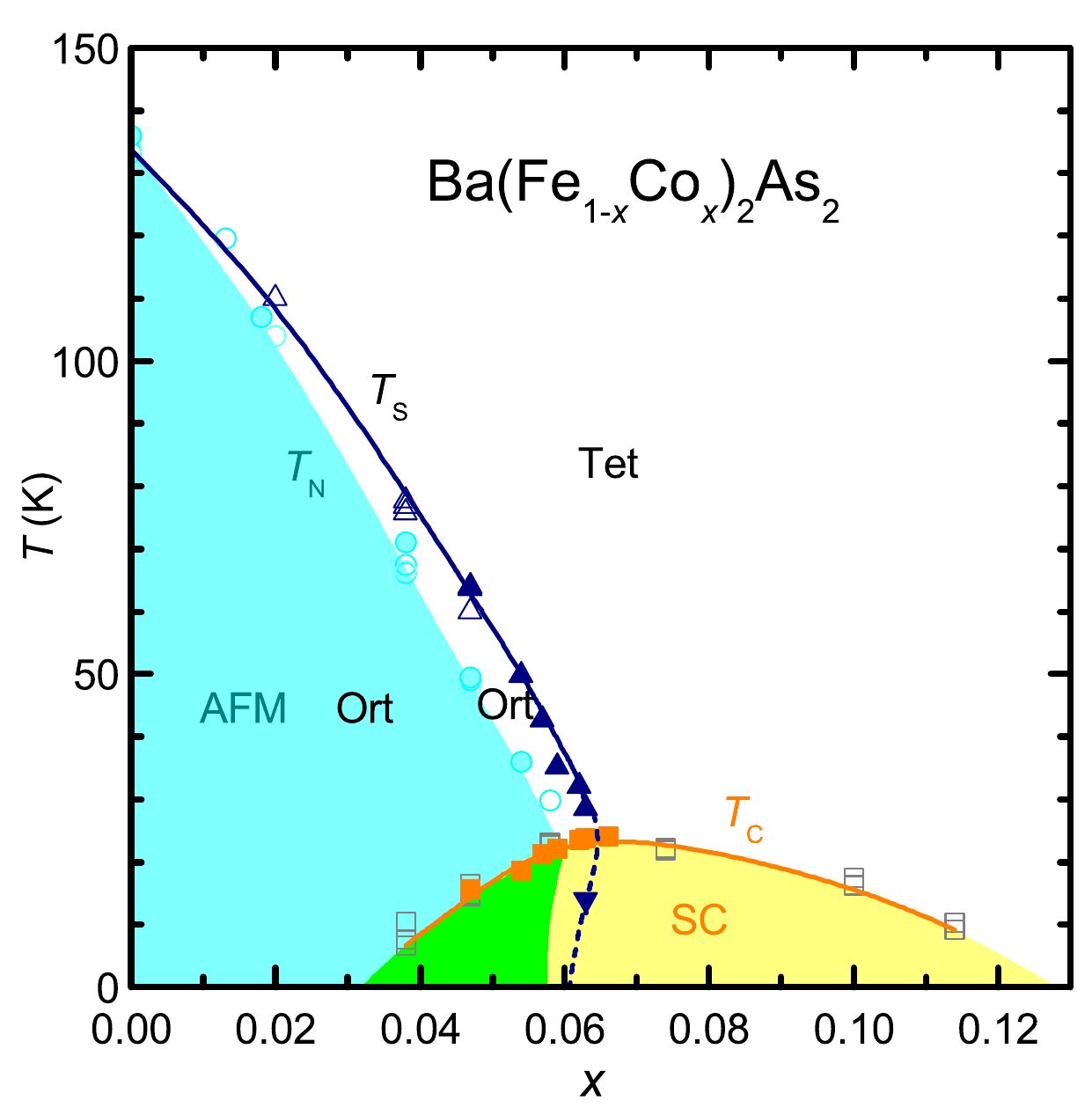}
\caption{Phase diagram of cobalt doped BaFe$_2$As$_2$ \cite{nandi}.  T$_s$ denotes
the structural transition, T$_N$ the antiferromagnetic one.}
\label{FeAs-phase}
\end{figure}

Upon doping, the magnetic and structural phase transitions are suppressed, and then superconductivity emerges (Fig.~\ref{FeAs-phase}).
Many of the materials indicate a significant range of dopings where the magnetic and superconducting orders co-exist.  As in cuprates,
a pronounced spin resonance is seen in the superconducting state, as well as a spin gap \cite{FeAs-res}.  Unlike the cuprates, where
the resonance appears to be a purely triplet excitation (as in cuprates, the pnictides appear to be singlet superconductors), there is evidence
that the resonance instead may be a doublet \cite{doublet}.  This may be due to the strong anisotropy of the magnetism, where the spin direction tends to
be locked to the iron layers.

Given the nature of the Fermi surface, it did not take long for a theory to emerge that suggested the existence of so-called $s_{\pm}$ pairing \cite{mazin-s}.
This state is a two band generalization of d-wave pairing, where the Fermi surfaces at $\Gamma$ have an opposite sign for the order parameter than the
surfaces at $M$.  The advantage of this state is that it has the required change of sign upon translation by the magnetic ordering vector $Q$
(necessary to obtain a solution in the BCS gap equation for magnetic mediated pairing), yet avoids having nodes, which typically cost energy.
This state is consistent
with subsequent photoemission studies \cite{ding-EPL} which indicated relatively isotropic gaps on the Fermi surface (Fig.~\ref{FeAs-PES}).
To date, though, there have only
been hints that such a state exists based on phase sensitive measurements \cite{tsui}, though it is certainly consistent with the observation of
a spin resonance \cite{FeAs-res} which implies a sign change of the order parameter under translation by $Q$.

\begin{figure}
\includegraphics[width=2.5in]{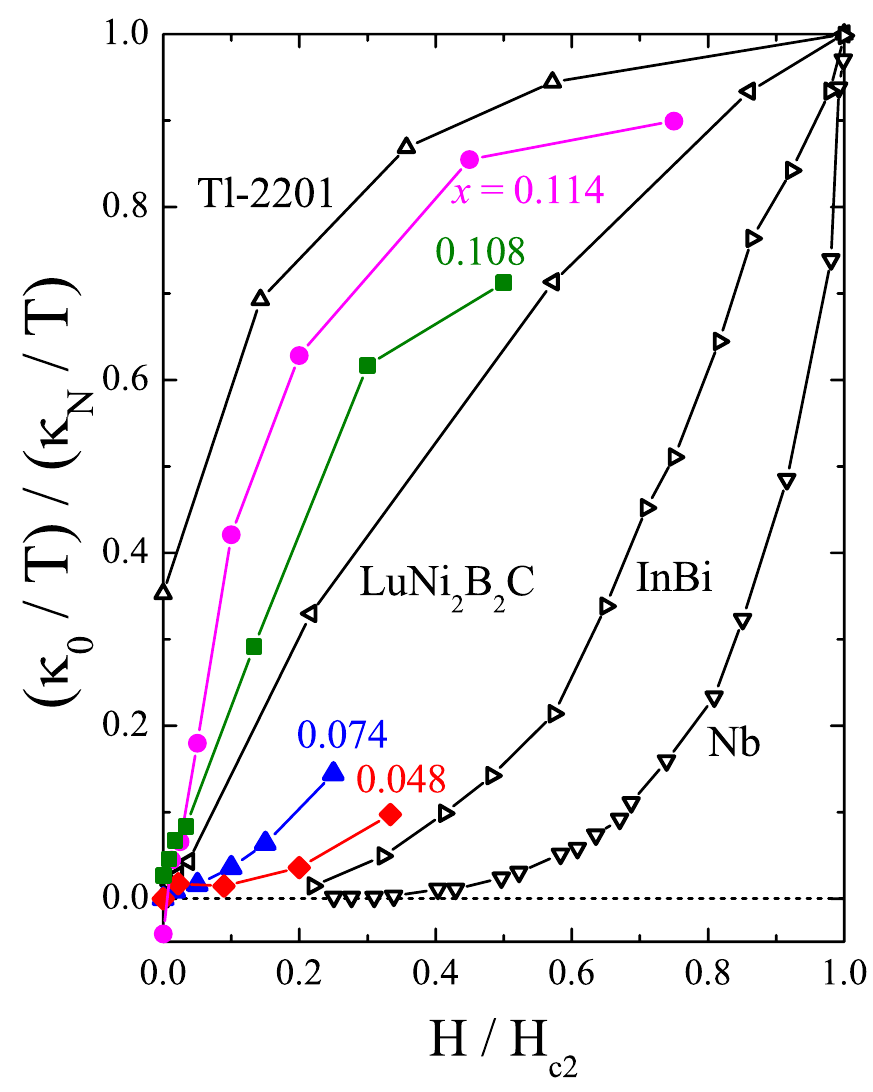}
\caption{Magnetic field dependence of the thermal conductivity of cobalt doped BaFe$_2$As$_2$ \cite{louis-FeAs}.
Note the evolution from s-wave like behavior (as in Nb) to d-wave like behavior
(as in the cuprate Tl2201) as the doping increases.}
\label{FeAs-K}
\end{figure}

Since then, a rich variety of information has become available from such probes as NMR, penetration depth, specific heat, and thermal conductivity.
Particularly telling has been the magnetic field dependence of the thermal conductivity  which indicates an evolution from a nodeless gap
to a gap with nodes as the doping increases \cite{louis-FeAs} (Fig.~\ref{FeAs-K}).  Other measurements indicate nodes or not depending on the material.
So far, there
has been little evidence for nodes from photoemission, though it should be remarked that pnictides exhibit substantial c-axis dispersion, as evidenced
by quantum oscillation studies \cite{FeAs-QO-2}, which means care should be taken with interpretations based on surface sensitive measurements.
The strive to address the gap anisotropy question by techniques with better energy resolution has propelled studies using Fourier transformed 
STM, which although also surface sensitive, allows the mapping of spanning vectors across the Fermi surface via quasiparticle interference arising 
from impurity scattering.  The most recent study indicates significant gap anisotropy on the $\Gamma$ centered hole surfaces
in LiFeAs \cite{STM-LiFeAs} but no nodes, which is consistent with photoemission studies \cite{ding-LiFeAs}.
An additional advantage of these FT-STM studies is that they can give information on the phase of the order parameter from the magnetic field evolution 
of the quasiparticle interference pattern, which is consistent with an $s_{\pm}$ state \cite{hanaguri}.

STM studies also reveal that dopants tend to dimerize \cite{davis-Co}, which may be related to the nematicity.  As this topic has gained much attention 
of late, it is worth going into in more detail.  Transport studies of the pnictides indicate that upon application of uniaxial pressure in the planes, a significant resistivity
anisotropy develops above the structural transition temperature \cite{nematic}.  This effect gets particularly pronounced at dopings where the 
structural transition begins to be suppressed.
To delve further, it should be remarked that unlike cuprates, in pnictides, band theory predicts that all five of the iron d bands are present in the vicinity 
of the Fermi energy, and these bands are well separated in energy from the p states of the ligands \cite{elec-FeAs}.  Nematicity is equivalent to
breaking the degeneracy between the $x$ and $y$ directions, which would imply a breaking of the degeneracy of the iron $d_{xz}$ and $d_{yz}$
orbitals.  This has now been observed by photoemission, where the polarization dependence of the data allows one to differentiate different d 
orbitals \cite{shen-EPL}.  The most likely explanation of this effect is local orthorhombic order that persists above the structural transition.  Since the
effect seems stronger than what one would anticipate due to the structure (as in the cuprates), the speculation is that it is either due to orbital ordering
as occurs in other transition metal oxides \cite{philip} or `spin' nematicity \cite{chub}.  The latter seems more likely, in that the magnetism and orthorhombicity
seem to be intimately related based on the observed phase diagram (Fig.~\ref{FeAs-phase}).  Even band theory studies reveal that the effective
exchange constants in the magnetic phase (where one perturbs about the ordered phase) are strong and antiferromagnetic along one in-plane direction
and weak and ferromagnetic along the other, consistent with the observed stripe order \cite{pickett2,yildirim}.  This pronounced anisotropy is also
evident in the spin wave dispersions in the undoped case \cite{dai-SW}.  If a `spin' nematic picture is correct, this is a further testament that
magnetic correlations may be responsible for the pairing.

In that context, there have been many attempts to calculate the pairing microscopically.  Early studies seem to rule out an electron-phonon
mechanism \cite{FeAs-ep}.  Most studies have not unsurprisingly indicated $s_{\pm}$ pairing due to magnetic correlations, with some of the
more unbiased studies based on the functional renormalization group (FRG) \cite{FeAs-FRG}.  Pairing due to antiferromagnetic spin fluctuations is on
more solid ground theory-wise than in the cupratres due to the somewhat weaker correlations in the pnictides.  Of course, the on-site $U$ is large for iron,
but unlike cuprates, the electrons can somewhat avoid one another by hopping to different d orbitals, leading to a smaller effective $U$ \cite{anisomov}.
And because of the presence of multiple $d$ orbitals, Hunds rule exchange plays a dominant role as compared to the superexchange $J$ of
the cuprates \cite{hunds}.  In fact, these findings have led to the quip that pnictides have freed us from the tyranny of Mott physics \cite{lee-quote},
though there are many in the field that would disagree.  Still, RPA, FRG, and DMFT studies are in broad agreement concerning the physics
of the superconducting state \cite{FeAs-theory}, though there is still some disagreement on how correlated the electrons are and the role
of Fermi surface nesting, as well as the doping dependence of the gap symmetry and
the importance of spin-lattice coupling in regards to the pairing.  Given the wealth of information on how the magnetic correlations, electronic structure,
and pairing evolve as a function of doping, it is anticipated that a well accepted theory will emerge in the near future.  What is particularly attractive
about these materials is that in many cases, the full doping range can be accessed (for instance, BaFe$_2$As$_2$ - KFe$_2$As$_2$)
as opposed to the cuprates where superconductivity only exists over a relatively narrow doping range of 20\%.

Another attractive feature of pnictides is that in many cases, they exhibit a full energy gap, and also have a relatively weak anisotropy (particularly
evident in the directional dependence of the upper critical field).  This means that they have the potential of being more technologically relevant
than the cuprates, at least in a certain temperature range, though the pnictides unfortunately show the same `crashing' of the critical current with 
grain boundary misalignment as occurs in the cuprates \cite{grain}.  This again may be related to the non-trivial phases associated with $s_{\pm}$ pairing.

\section{Other Classes}

Space prohibits a detailed discussed of other classes of unconventional superconductors, though a few of them are definitely worth mentioning here.

A number of transition metal oxides have the same crystal structure as La$_2$CuO$_4$.  Of particular interest is Sr$_2$RuO$_4$, which exhibits
superconductivity at 1.5 K \cite{maeno}.  The superconductivity in this case appears to be triplet in nature \cite{sigrist-SRO}, though there is still
much debate on the nature of the order parameter.  Phase sensitive measurements are consistent with a sign change of the order parameter
when comparing opposite faces \cite{mao} and there is evidence as well for time reversal symmetry breaking below T$_c$ \cite{SRO-muSR}
which has been taken as support for a $(k_x \pm ik_y){\hat z}$ pair state.  Several complications exist that indicate the order parameter may not be as simple as
this.  First, three d bands comprise the Fermi surface, as evidenced by quantum oscillation \cite{SRO-QO} and ARPES studies \cite{SRO-PES},
all sitting near the Brillouin zone boundary, so a simple  $k_x \pm ik_y$ form (based on an expansion near $\Gamma$) is unlikely.
Moreover, if triplet, the $d$ vector structure of the gap has still to be verified, given technical difficulties with obtaining NMR data at low enough magnetic fields,
along with the potential issue of field re-orientation of the $d$ vector \cite{SRO-NMR}.  In that connection, the role of 
spin-orbit coupling in determining the pair state has yet to be fully elucidated.
A detailed discussion of this fascinating material is beyond the scope of this chapter, so the reader is referred to a number of excellent
reviews for more information than provided here \cite{SRO-NMR,SRO-reviews}. Certainly, the excitement surrounding strontium ruthenate is not only that 
it may be the realization of p-wave pairing that had been long sought in materials
such as palladium, but because of the chiral nature of the proposed order parameter, it could potentially be exploited for topological
quantum computing \cite{sarma-RMP}.  In that connection, recent experiments are consistent with the existence of half quantum vortices \cite{half}.

It was Phil Anderson's great insight in realizing that even in the presence of strong spin-orbit coupling, one could still use parity to classify pair states,
and connect this parity with a `singlet' and a `triplet' in the degenerate space of $k$, $Pk$, $Tk$, and $PTk$, where $P$ is the parity operator and $T$ the
time reversal one \cite{pwa84}.  This is not only relevant for heavy fermion superconductors, but potentially for strontium ruthenate as well as
mentioned above.  But what if parity is broken?  This is an old problem going back to magnetic superconductors, since the magnetic structures
of antiferromagnets typically break parity symmetry (though they preserve the product of $P$ and $T$).  As mentioned earlier, this should lead to a gap
function which is a mixture of a primary even parity component associated with center of mass momentum zero, and a secondary odd parity component
with center of mass momentum $Q$ (with $Q$ the magnetic ordering vector) \cite{fenton}.  Of course, there is the simpler case where the crystal structure
itself breaks parity.  In that case, one would assume that even if the primary component of the order parameter was an even parity singlet, an odd
parity component could be mixed in.  In the past decade, a number of such non-centrosymmetric superconductors have been discovered \cite{NC-SC}.
Space prohibits a detailed discussion of this class of materials, though the interesting ones typically involve ions where spin-orbit coupling
is strong.  One of these materials is Li$_2$X$_3$B with $X$ either Pd or Pt.   The Pd case looks singlet in nature and the Pt case
triplet \cite{LiPdB}, though detailed studies of the pairing symmetry as done in the cuprates have yet to be performed.  Given their nature, though,
non-trivial topological properties of these materials might be realized \cite{NC-topo}.

This brings us to topological superconductors, a topic of much current interest \cite{topo}.  Topological insulators are variants of normal band insulators
where a non-trivial Berry phase exists \cite{kane}.  The effect of this non-trivial phase is the presence of gapless surface states.  These have
been observed in such systems as Bi$_2$Se$_3$ \cite{bi2se3}.  The connection with superconductivity is two fold.  First, the prediction that if such
materials could be made superconducting, they might exhibit p-wave pairing \cite{fu}.  Cu doped  Bi$_2$Se$_3$ is superconducting \cite{cu-bi2se3},
but the jury is still out on the nature of its pairing, though p-wave pairing seems unlikely (the latest tunneling measurements indicate an isotropic energy
gap without any in-gap states \cite{BiSe-tun}).  The other connection is that if a topological insulator
is brought into contact with an s-wave superconductor, zero energy bound states can be induced that behave like Majorana 
fermions \cite{fu-kane,luchtyn,oreg},
that is, particles that are their own anti-particles.  Such fermions could in principle be manipulated for the purpose of topological quantum
computing \cite{topo-majo}.
Zero energy bound states have indeed been seen in such hybrid systems \cite{leo}, but a unique identification of the bound states as Majoranas is a
subject of much study and debate \cite{majorana}.  Certainly, this is a very active field which is anticipated to yield significant results in the next few years.
In that context, p-wave `spinless' supercurrents have been induced via the proximity effect in half metallic magnets like CrO$_2$ where the carriers are fully
spin polarized \cite{keizer}.

Finally, a number of strong coupling superconductors that are likely s-wave have been identified over the years \cite{pickett-01}.
MgB$_2$ has a particularly high T$_c$ due to coupling to a particular high energy phonon mode, though it is essentially a conventional superconductor.  Shortly after
the discovery of cuprates, superconductivity at 30 K was discovered in the perovskite Ba$_{1-x}$K$_x$BiO$_3$ \cite{BKBO} following earlier work
on the lower T$_c$ lead analogue.  A relatively high T$_c$ of 25 K is also seen in a layered sodium doped HfNCl material \cite{HfNCl}.  Recent
theoretical work on these materials is consistent with electron-phonon pairing that is enhanced by strong electron correlations \cite{yin}.
Based on this, there have been recent predictions of related materials that might be superconducting \cite{yin2}.
The connections of these materials with unconventional superconductors like the cuprates has been a subject of much speculation.

\section{Experimental Trends}

A general observation from the previous sections is that the phase diagrams of unconventional superconductors look remarkably similar in many aspects.
Typically, superconductivity is obtained by suppressing some other ordered phase, such as antiferromagnetism.  This in turns links these materials
to more conventional superconductors such as transition metal dichalcogenides, which typically become superconducting once charge density
wave order is suppressed \cite{cava}.  Of course, different orders competing for gapping the Fermi surface is an old observation which is relevant to A15 
superconductors
as well \cite{bilbro}, but the presence of a quantum critical point in the phase diagram associated with the competing order
which is typically buried under the superconducting `dome'
is a potentially universal observation that cannot be ignored.  This is particularly relevant for those materials which exhibit quantum critical behavior
for temperatures above this critical point, which again seems universal to heavy fermions, cuprates, and pnictides.  Besides the intriguing prospect of a pairing
instability emerging from a non-Fermi liquid normal state, the idea that quantum critical fluctuations are the pairing `glue' is an appealing concept.
This is why there has been so much debate on the nature of the pseudogap phase \cite{AP}, since it is thought by many that the quantum critical fluctuations
associated with the suppression of this state could be the origin of cuprate superconductivity.  Whether a universal theory of unconventional
superconductivity is possible based on these ideas, and whether this is a large enough `umbrella' to capture much of the thinking on these materials
remains to be seen.  Although RVB ideas seem anathema to this line of approach, in some sense, superconductivity from this theory emerges from
the suppression of a Mott insulating phase, and so could potentially be captured in this framework as well.  Certainly, DMFT simulations are in support
of this picture.  We will certainly know more along these lines once we have `nailed' the phase diagram from experiment for various materials and
are able to properly correlate them from one class of materials to the next.

The other interesting trend is that certain crystal structures, such as the ThCr$_2$Si$_2$ one, seem to be amenable to superconductivity.  Whether this
is an accident or something profound remains to be seen.  Certainly, as discussed above for the UX$_3$ series of compounds, it is interesting that
the cubic ones do one thing, but it is the hexagonal variants that exhibit either superconductivity or novel quadrupolar order.  Again, as our database
of unconventional superconductors grows, the role that the crystal structure plays should become more evident.

\section{Theoretical Trends}

In some sense, we were very fortunate for conventional superconductivity that a unique strong-coupling theory emerged so rapidly after the BCS theory
was first proposed.  This was to a large part due to the Migdal theorem.  In strong coupling electron-phonon systems that are outside of this framework,
for instance those
exhibiting polaronic effects, there has yet to emerge a similarly robust calculational scheme.  For electron-electron pairing, though, all bets are off.
Migdal-like approximations have been invoked that attempt to exploit the separation of energy scales between collective and single particle degrees
of freedom \cite{abanov}, but the efficacy of this approach has yet to be demonstrated to everyone's satisfaction.  This had led to a host of approaches
that have been proposed - fluctuation exchange approximation \cite{FLEX}, functional renormalization group \cite{FRG}, two particle 
self-consistency theory \cite{tremblay}, large $N$ approaches \cite{large N}, and dynamical mean field theory in its various cluster
versions \cite{cluster-DMFT}.  Each of these methods has their pros and cons.  Unlike electron-phonon theories, where we know that there is
an attractive interaction (negative electrons, positive ions), in electron-electron theories where the bare Coulomb interaction is of course repulsive,
the `attractive' component is typically of an induced nature, making its evaluation (and even its sign!) a highly non-trivial process.  Anderson has
advocated that in RVB theories, a `glue' does not exist \cite{glue}, that is, there is no induced interaction, with $J$ itself being the pairing interaction.
On the other hand, most implementations of this theory have been done at a mean field level.  Variational Monte Carlo simulations have been done
which give very intriguing results in support of the basic conjectures of this theory \cite{randeria}, but these simulations are biased by nature.
Gauge theory approaches have been advocated which brings in dynamics \cite{lee-RMP}, but whether this represents a systematic approximation
has been questioned.  Certainly, recent developments in quantum Monte Carlo and DMRG methods are promising in regards to giving unbiased
results.  Coupled with other approaches, for instance cluster DMFT, there is some promise that results will emerge that will generally be
accepted by the community.

\begin{figure}
\includegraphics[width=2.25in]{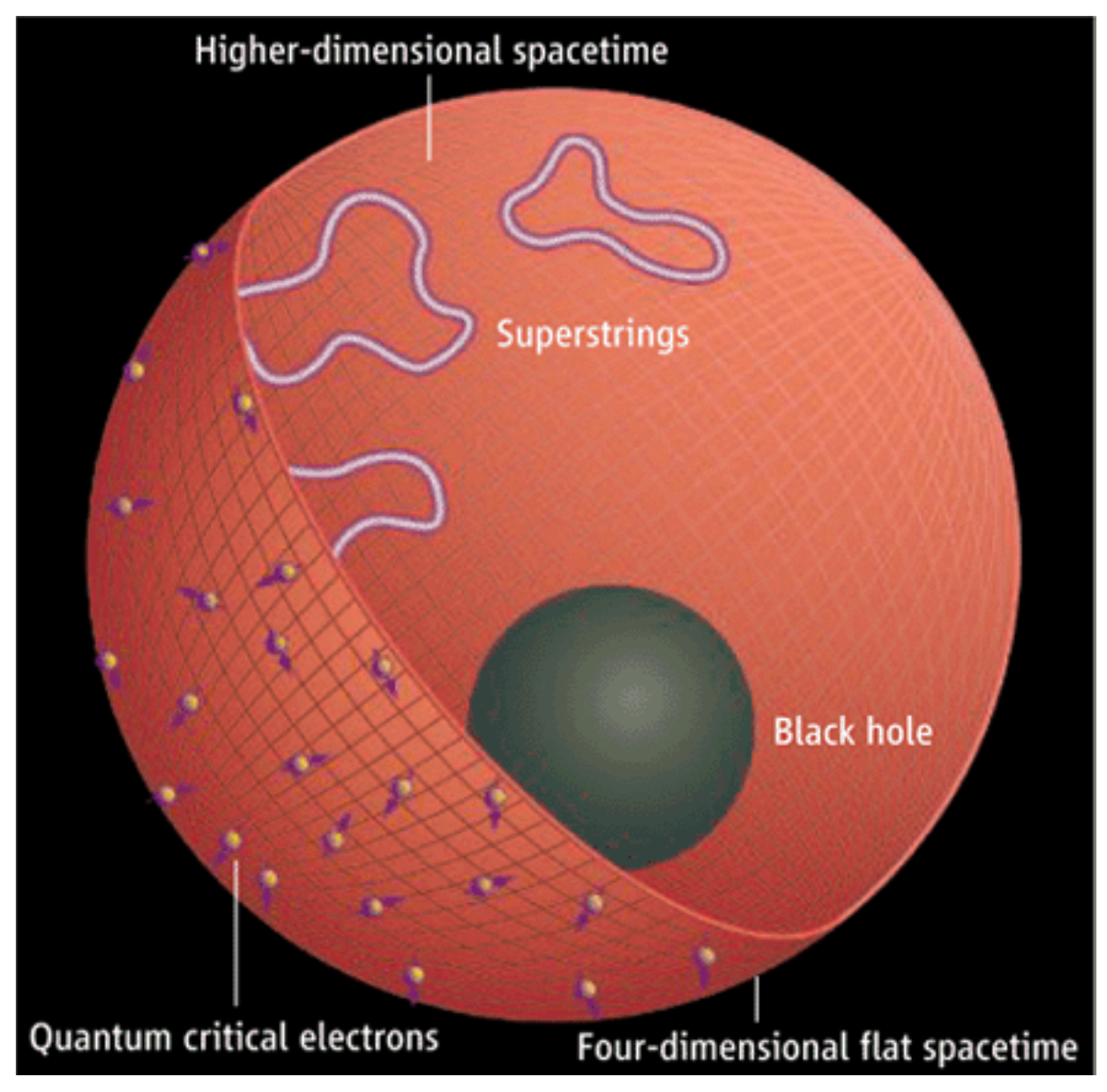}
\caption{A schematic representing the anti de Sitter (AdS) - conformal field theory (CFT) duality \cite{hartnoll}.  Quantum
critical electrons on the boundary of an AdS space-time are mapped to a weak coupling gravity dual
in the interior.  The black hole sets the charge density and temperature of the theory.}
\label{holo}
\end{figure}

However, the potentially non perturbative nature of this problem has led to the realization that new approaches might be needed to ultimately
solve the problem of unconventional superconductivity.  In that sense, the recent attention given to holographic theories is worth commenting on.
Strong coupling gauge theories in the context of particle physics has been notoriously difficult to come to grips with.  This led to the Maldacena
conjecture \cite{maldacena}.  This conjecture is based on mapping a strong coupling gauge theory that exists on the boundary of a fictitious
space-time to a weak-coupling gravitational theory in the bulk (Fig.~\ref{holo}).  The space-time in question is anti de Sitter (AdS) space, which is 
hyperbolic in nature.
The extra coordinate in this space-time can be thought of as the coordinate along a renormalization group flow, where one flows
from ultraviolet at the boundary to infrared in the interior.  To represent a charged system at non-zero temperature, one can simply put a black
hole in this space.  In the condensed matter context in two dimensions, one typically flows from an AdS$_4$ geometry near the boundary to
an AdS$_2$ times $R_2$ one near the black hole horizon \cite{liu}.  The net result is local quantum criticality, since the spatial $R_2$ part has
essentially decoupled (AdS$_2$ being dual to CFT$_1$, a conformal field theory in time).  In that sense, it is similar to the Kondo problem, which 
is local in space and critical in time \cite{piers}.  By changing various
parameters of the theory, one can tune from a Fermi liquid to a marginal Fermi liquid to a non Fermi liquid \cite{liu}.  Introduction of a scalar field can
lead to Bose condensation, thus `holographic' superconductivity \cite{holo-SC}.  Given certain stability conditions of a scalar field in an anti de Sitter
spacetime near a black hole, this condensation is dependent on the black hole horizon, and thus the temperature, therefore one can get a second 
order phase transition as a function of temperature just as in a real superconductor.

Although this approach is truly non perturbative in nature, for most applications, the theory is essentially at the Ginzburg-Landau level.  That is,
one assumes a scalar field.  Since it is a scalar, it should correspond to some charge $2e$ field, but since the theory does not explicitly invoke
pairing, extra terms have to be added to the action to describe the coupling of fermions to the scalar field (i.e, to generate a Bogoliubov dispersion).
One neat aspect is that this coupling is dependent on the geometry of the problem (which changes from AdS$_2$ times R$_2$ to AdS$_4$
once the scalar field condenses), and thus one can get the same `peak-dip-hump' structure observed by photoemission for the antinodal spectral 
function of the cuprates, basically since the fermion
damping is gapped outside an $\omega-k$ `light cone' determined by the geometry \cite{AdS-pdh}.  On the other hand, since these are in essence
phenomenological considerations, this does not bring real understanding to the problem, since there are a variety of physical effects that can
give rise to such a lineshape (including trivial effects like bilayer splitting).  Perhaps in the end, these approaches will help to resolve issues
connected to the gauge theory approaches used for both the Kondo problem \cite{piers} and RVB theories \cite{lee-RMP}.
These theories are non-trivial since the gauge fields
involved are associated with constraints, and thus differ in fundamental ways from the gauge theories quantum field theorists typically study \cite{nayak}.

\section{The Future}

As Yogi Berra supposedly quipped, ``It's tough to make predictions, especially about the future".  Still, given developments in superconductivity over 
the past several decades, it is worth giving it a shot.  First, given the number of new classes of materials discovered in the past thirty years or so,
many of them not touched on here \cite{mandrus}, it is pretty certain that new classes of unconventional superconductors will be
discovered in the near future.  And given the fact that a number of these classes have high T$_c$, it is pretty certain as well that new high T$_c$
materials are in the offing, though it is unclear
whether we will ever beat the cuprates in T$_c$ (at least, under earth like conditions - witness the possibility of ultra high T$_c$ in metallic
hydrogen \cite{ashcroft}).  Most of these discoveries will certainly be serendipitous, though there is hope with the development of
layer by layer synthesis, for instance by molecular beam epitaxy, that one might `design' superconductors \cite{bozovic}.
But as in the old days, a lot of the discoveries will be by people following their nose, as Muller did for cuprates, and Hosono for pnictides.
In that context, there have been recent speculations that doped iridium oxides will be superconducting because of the large value of
the exchange integral \cite{senthil}, but to date, this prediction has yet to be verified.

The other matter to comment on is theory.  As numerical techniques continue to improve, more and more will be known about non-trivial
many body theories from `exact' techniques like quantum Monte Carlo, and DMRG and related approaches.
Moreover, the evolution of cluster DMFT into a predictive
tool for superconductivity \cite{gabe-EPL} is a very welcome development.  In the end, though, it will take rigorous solutions to convince
the skeptics that the results are not due to some sort of calculational bias (i.e., by making the assumption that the model actually has a superconducting
ground state to being with).  After all, there are few exact theories we know of superconductivity outside of the electron-phonon model.  The Kohn-Luttinger
theory comes to mind \cite{KL}, though this predicted instability of the normal state only occurs at very low temperatures.

Finally, it could well be that `materials genome' databases might yield new predictions via data mining \cite{klint}, assuming one is using valid 
search criteria.  That is, that one is not operating in the GIGO (garbage in, garbage out) mode.  Certainly, the phase space of materials to
explore is astronomical, and it will definitely take a lot of imagination, both from theory and experiment, to explore its infinite richness.

\begin{acknowledgments}
This is an expansion of an even briefer review written by the author for Science \cite{science}.  The author would like to thank Dr.~Jelena Stajic for providing
the opportunity to write that paper, as well as Prof.~John Ketterson for the opportunity of expanding it for this book.
This work was supported by the Center for Emergent Superconductivity, an Energy Frontier Research
Center funded by the U.~S.~Dept.~of Energy, Office of Science, under Award No.~DE-AC0298CH1088.
The author would like to thank his colleagues in the CES that provided much of the inspiration presented here.
\end{acknowledgments}

\end{document}